\documentclass[useAMS,usenatbib]{mn2e}
\usepackage{graphicx}
\usepackage{latexsym}
\usepackage{xspace}
\usepackage{amsmath}

\title[CO-dark gas in the Milky Way]{CO-dark gas and molecular filaments in Milky Way type galaxies}
\author[Smith et~al.]{Rowan J. Smith$^{1}$\thanks{Email: rowan@uni-heidelberg.de}, Simon C. O. Glover$^{1}$, Paul C. Clark$^{1}$,   
\newauthor Ralf S. Klessen$^{1}$,  and Volker Springel$^{2,3}$ \\
$^{1}$Zentrum f\"ur Astronomie der Universit\"at Heidelberg, Institut f\"ur Theoretische Astrophysik, Albert-Ueberle-Str. 2, \\ 
69120 Heidelberg, Germany \\
$^{2}$Heidelberger Institut f\"ur Theoretische Studien, Schloss-Wolfsbrunnenweg 35, 69118 Heidelberg, Germany \\
$^{3}$Zentrum f\"ur Astronomie der Universit\"at Heidelberg, Astronomisches Recheninstitut, M\"onchhofstr.\ 12--14, \\
69120, Heidelberg, Germany
 }

\begin{document}

\pagerange{\pageref{firstpage}--\pageref{lastpage}} \pubyear{2010}

\maketitle

\label{firstpage}

\def\mnras{MNRAS}
\def\apj{ApJ}
\def\aap{A\&A}
\def\apjl{ApJL}
\def\apjs{ApJS}
\def\bain{BAIN}
\def\pasp{PASP}
\def\araa{ARA\&A}
\def\ga{\sim}
\def\nat{Nature}
\def\aj{AJ}
\def\pasj{PASJ}


\newcommand{\eq}{Equation }
\newcommand{\fig}{Figure }
\newcommand{\tab}{Table }
\newcommand{\msun}{\,M$_{\odot}$ }
\newcommand{\gcmc}{\,g\,cm$^{-3}$}
\newcommand{\kms}{\,km\,s$^{-1}$\xspace}
\newcommand{\gcms}{\,g\,cm$^{-2}$\xspace}
\newcommand{\E}{\times 10}
\newcommand{\cms}{\,cm$^{-2}$\xspace}
\newcommand{\cmc}{\,cm$^{-3}$\xspace}
\newcommand{\sd}{\,M$_{\odot}$ pc$^{-2}$\xspace}
\renewcommand{\deg}{\ensuremath{^{\circ}}\xspace}

\newcommand{\arepo}{\textsc{arepo}\xspace}
\newcommand{\sph}{SPH }
\newcommand{\treecol}{{\sc TreeCol}\xspace}
\newcommand{\healpix}{\textsc{healpix}\xspace}
\newcommand{\gadget}{\textsc{gadget} 2\xspace}

\newcommand{\NHm}{$N_{\rm H_2}$\xspace}
\newcommand{\NCO}{$N_{\rm CO}$\xspace}
\newcommand{\WCO}{$W_{\rm CO}$\xspace}
\newcommand{\ZCO}{$Z_{\rm CO}$\xspace}
\newcommand{\XCO}{$X_{\rm CO}$\xspace}


\begin{abstract}
We use the moving mesh code \arepo coupled to a time-dependent chemical network to investigate molecular gas in simulated spiral galaxies that is not traced by CO emission. We calculate H$_{2}$ and CO column densities, and estimate the CO emission and CO-H$_2$ conversion factor. We find that in conditions akin to those in the local interstellar medium, around 42\% of the total molecular mass should be in CO-dark regions, in reasonable agreement with observational estimates. This fraction is almost insensitive to the CO integrated intensity threshold used to discriminate between CO-bright and CO-dark gas. The CO-dark molecular gas primarily resides in extremely long ($>100$ pc) filaments that are stretched between spiral arms by galactic shear. Only the centres of these filaments are bright in CO, suggesting that filamentary molecular clouds observed in the Milky Way may only be small parts of much larger structures. The CO-dark molecular gas mainly exists in a partially molecular phase which accounts for a significant fraction of the total disc mass budget. The dark gas fraction is higher in simulations with higher ambient UV fields or lower surface densities, implying that external galaxies with these conditions might have a greater proportion of dark gas. 
\end{abstract}


\section{Introduction} \label{intro}
The nature and distribution of molecular gas in the interstellar medium (ISM) is of great interest to astrophysicists because all local star formation occurs within clouds of molecular gas. Unfortunately, the main constituent of this gas, molecular hydrogen (H$_{2}$), is extremely difficult to observe directly, owing to the large energy separation between its rotational levels, as well as its absence of a dipole moment. Most studies of H$_{2}$ in the ISM therefore involve observing a tracer species, most commonly carbon monoxide (CO), and then inferring the properties of the molecular gas as a whole from the behaviour of this tracer. A crucial assumption here is that the tracer does indeed trace all of the H$_{2}$. However, there is growing observational evidence that this is not the case for CO, and that there is a component of the dense ISM that is rich in H$_{2}$ but that produces little or no CO emission \citep{Tielens85,vanDishoeck88}.

Evidence for the existence of this component, termed ``dark molecular gas'' by \citet{Wolfire10}, comes from several different sources. Gamma-ray observations of the local ISM probe the total hydrogen column density, regardless of whether the hydrogen is in the form of H$^{+}$, H or H$_{2}$. Comparisons between maps of  the diffuse gamma-ray emission and maps of the H{\sc i} 21~cm emission and CO emission show that there is a considerable amount of hydrogen surrounding molecular clouds that is not well traced by either form of emission, and that has therefore been interpreted as being CO-dark molecular gas \citep{Grenier05,Abdo10}. Measurements of total gas column densities using dust extinction or dust emission allow a similar comparison to be made and also show evidence for a significant fraction of CO-dark molecular material \citep[see e.g.][]{RomanDuval10,Ade11,Paradis12}. Further corroboration comes from \citet{Allen2012}, who find bright OH emission that is uncorrelated with atomic hydrogen or CO emission, and from \citet{PinedaJ13}, who show that a significant CO-dark molecular gas component is needed in order to explain the distribution of [C{\sc ii}] emission in the Galaxy. Finally, this ``dark'' molecular component can be probed directly by UV absorption line measurements \citep[see e.g.][]{Burgh07,Sonnentrucker07,Sheffer08}, albeit only along lines-of-sight where there happens to be a UV-bright background source. These UV absorption line studies confirm that significant H$_{2}$ can be present in regions without much CO, but do not allow this gas to be mapped, or its total mass to be determined.

Numerical simulations can provide valuable assistance for interpreting this observational data, and allow us to better understand how ``dark'' H$_{2}$ is distributed in the ISM. For example, \citet{Wolfire10} use photodissociation region (PDR) models of a series of clouds to quantify the amount of dark H$_{2}$ that one would expect to find in their envelopes, finding values of around 30--40\% of the total H$_{2}$ content of the clouds, with little dependence on cloud properties other than metallicity and mean extinction. Although illuminating, this kind of 1D PDR model must inevitably make use of a highly approximate treatment of cloud structure. Moreover, the \citet{Wolfire10} models also assume chemical steady-state, and hence do not allow one to examine the effects of the dynamical history of the gas on its H$_{2}$ content, even though 3D dynamical models suggest that this may be highly significant \citep{Glover07a,Glover07b,Dobbs08a,Glover10}.

Ideally, what one would like to do is to model the dynamical evolution of the ISM together with its chemical evolution in a self-consistent fashion, in order to better quantify how much dark molecular gas may be found around CO-bright molecular clouds, and also how much is distributed in the form of diffuse clouds with high H$_{2}$ fractions but little or no CO. However, to do this requires one to model both the H$_{2}$ and CO chemistry within a high dynamical range 3D simulation of the evolution of the ISM. Until very recently, this has not been possible. A number of groups have performed large-scale dynamical simulations of the ISM that account for H$_{2}$ formation \citep[see e.g.][]{Dobbs08a,Gnedin09,Christensen12}, but these models typically have insufficient resolution to allow one to study individual GMCs in detail, and they are also missing any treatment of the CO chemistry. On the other hand, small-scale models have for some time been able to model both H$_{2}$ and CO formation in the turbulent ISM \citep[see e.g.][]{Glover10,Glover12b}, but have not had the dynamical range necessary to model cloud formation self-consistently, or have started from highly artificial initial conditions 
\citep{Clark12}.

In the present paper, we present initial results from  a study that attempts to combine a large-scale 3D model of the dynamically evolving ISM with a detailed treatment of the small-scale cooling physics and chemistry. Our goals in this paper are to quantify how much CO-dark gas one would expect to find in the ISM, to explore how the fraction of molecular gas that is CO-dark varies as we change the mean surface density of the galaxy or the strength of its interstellar radiation field, and to better understand the spatial distribution of the CO-dark molecular gas. In Section \ref{methods}, we introduce our numerical methods and discuss the initial conditions that we use for our simulations. In Section \ref{MW}, we analyse in depth the dark gas fraction, morphology, molecular abundance and estimated CO emission for our fiducial case based on the Milky Way galaxy. In Section \ref{comparison}, we then compare to three simulations with different mean surface densities and ambient radiations fields. Finally, in Section \ref{discussion} we discuss our results, and in Section \ref{conclusions} we give our conclusions.

\section{Methods} \label{methods}

\subsection{Hydrodynamical model} \label{hydro}
We perform our simulations using the moving mesh code \arepo \citep{Springel10}. \arepo is a novel hydrodynamical code that attempts to combine the strengths of grid-based Eulerian hydrodynamical codes and smoothed particle hydrodynamics (SPH). It solves the hydrodynamical equations on an unstructured mesh, defined as the Voronoi tessellation of a set of mesh-generating points that can move freely with the gas flow. The freedom of movement of the mesh-generating points allows the method to smoothly adjust its resolution to account for local clustering, in a similar fashion to SPH, while its use of a mesh allows high-order shock-capturing schemes to be used, as in modern adaptive mesh refinement (AMR) codes. The \arepo mesh is adaptable and can be refined to give improved mass resolution in regions of interest. This allows the study of problems with an extreme dynamic range, that are discontinuous, and involve fluid instabilities, all while imparting no preferred geometry on the problem. In addition, \arepo has been shown to perform much better than standard SPH for modelling highly multiphase flows \citep{Sijacki12} and subsonic turbulence \citep{Bauer12}. These properties make \arepo an ideal tool for studying the formation of dense molecular gas in galactic spirals. \arepo also has the capability to include magnetic fields \citep{Pakmor11} but in this first paper we restrict ourselves to the hydrodynamic case.

\subsection{Chemical model}
\label{chem}

The chemical evolution of the gas in our simulations is modelled using the hydrogen chemistry of \citet{Glover07a,Glover07b}, together with the highly simplified treatment of CO formation and destruction introduced in  \citet{Nelson97}. Our treatment of the hydrogen chemistry includes H$_2$ formation on grains, H$_{2}$ destruction by photo-dissociation, collisional dissociation and cosmic ray ionisation, collisional and cosmic ray ionisation of atomic hydrogen and H$^{+}$ recombination in the gas phase and on grain surfaces (see Table 1 of \citealt{Glover07a}). The evolution of the CO abundance is calculated using the assumption that the CO formation rate is limited by an initial radiative association step, and that the CO destruction rate is primarily due to photodissociation. Full details of the combined network, and a discussion of how it compares to other approaches in terms of accuracy and speed, are given in \citet{Glover12a}: the network used here is the same as the NL97 model in that paper.

In our fiducial `Milky Way' model and our `Low Density' model (see Section~\ref{init} below), we assume that the strength and spectral shape of the ultraviolet portion of the interstellar radiation field (ISRF) are the same as the values for the solar neighbourhood derived by \citet{Draine78}; note that this corresponds to a field strength of 1.7 times the field strength used by \citet{Habing68}. We also consider a `Strong Field' model, in which the radiation field strength is increased by a factor of ten compared with this fiducial value, and a `Low \& Weak' model, in which it is decreased by a factor of ten. 

To treat the attenuation of the ISRF due to H$_{2}$ self-shielding, CO self-shielding, the shielding of CO by H$_{2}$, and by dust absorption, we use the \treecol algorithm developed by \citet{Clark12b}. This algorithm computes a $4\pi$ steradian map of the dust extinction and H$_{2}$ and CO column densities surrounding each \arepo cell, using information from the same oct-tree structure that \arepo uses to evaluate gravitational interactions between cells. The resulting column density map is discretised onto $N_{\rm pix}$ equal-area pixels using the {\sc healpix} algorithm \citep{healpix}. In the simulations presented here, we set $N_{\rm pix} = 48$. To convert from H$_{2}$ and CO column densities into the corresponding shielding factors, we use shielding functions taken from \citet{Draine96} and \citet{Lee96}, respectively. These are pre-calculated by integrating over the adopted ISRF, and so they can be simply scaled with the radiation field strength in the simulations to find the relevant attenuation factor for a given column density.

The original version of the \treecol algorithm presented in \citet{Clark12b} considers all of the gas between the cell or particle of interest and the edge of the computational volume when computing the column density or the dust extinction. This is a reasonable choice if one is interested in modelling an isolated molecular cloud \citep[see e.g.][]{Glover12b}, but in the large-scale simulations presented here, it would lead to a substantial overestimate of the column densities along lines of sight passing through the mid-plane of the disc. To avoid this, we define a shielding length $L_{\rm sh} = 30 \: {\rm pc}$, and when calculating our column density and extinction maps, we include contributions only from gas located at a distance $L \leq L_{\rm sh}$ from the \arepo cell of interest. Our choice of a value for $30 \: {\rm pc}$ for $L_{\rm sh}$ is motivated by the fact that in the solar neighbourhood, the typical distance to the nearest O or B star is of order 30~pc \citep{Reed00,Maiz-Apellaniz01}.

Finally, we account for cosmic ray ionisation of H and H$_{2}$. In our fiducial `Milky Way' model, we adopt a rate $\zeta_{\rm H} = 3 \times 10^{-17} \: {\rm s^{-1}}$ for atomic hydrogen, and a rate twice the size of this for molecular hydrogen. In our other models, we scale the cosmic ray ionisation rates linearly with the strength of the ISRF, under the assumption that the same massive stars that power the UV field are also responsible, after their deaths, for generating the cosmic rays. We therefore adopt a cosmic ray ionisation rate in our `Strong Field' simulation that is ten times larger than in our `Milky Way' simulation, while in the `Low \& Weak' model, the rate is ten times smaller.

\subsection{Galactic potential}
\label{pot}
The final ingredient needed to study molecular cloud formation is a model for the galactic potential. For this we adopt the approach of \citet{Dobbs06} and impose an analytical potential representing a four-armed spiral upon a disc of gas without self-gravity. A logarithmic potential that produces a flat rotation curve of $v_0=220$ \kms \citep{Binney87} is combined with a potential for the outer halo from \citet{Caldwell87}. This is then perturbed by a four-armed spiral component from \citet{Cox02}. The spiral potential has a pitch angle $\alpha=15\deg$, and a pattern speed of $2\E^{-8}$ rad yr$^{-1}$, which corresponds to a co-rotation radius of 11 kpc. The gas is therefore orbiting faster than the spiral pattern in these simulations.

This scheme is of course an oversimplification that neglects stellar feedback and the self-gravity of the gas. The lack of stellar feedback means that we will have too many molecular clouds in the simulation compared to models with feedback \citep[c.f.][]{Hopkins12a}. However, this will to some extent be offset by the lack of gas self-gravity, which will generally decrease the number of large clouds formed. The absence of self-gravity is unlikely to be a major obstacle in forming H$_2$ and CO as the transition from atomic to molecular gas occurs at number densities of about 1-100 \cmc, where the contribution from self-gravity is still small compared to the global potential. Owing to the lack of stellar feedback, we advise treating the total values of the H$_2$ and CO masses in our simulation with some caution. However, we reproduce the morphology of a multiphase, filamentary, spiral galaxy well and therefore believe that we have an accurate picture of the \textit{ratio} of CO to H$_2$ across the disc. If anything, our lack of feedback will lead us to underestimate the amount of dark gas, as there will be more dense, fully-molecular clouds that are bright in CO emission than there would be if we were to include feedback.

\subsection{Initial Conditions} \label{init}

The initial setup of each simulation is a torus of thickness 200~pc, inner radius 5~kpc, and outer radius 10~kpc. The potential applied to the gas imparts a clockwise radial velocity of 220 \kms and causes the gas to concentrate in the disc mid-plane. We do not include the inner portion of the disc, both for reasons of computational efficiency, and also because the behaviour of gas in this region will be strongly influenced by interactions with the galactic bar, if one is present. In addition, we expect the interstellar radiation field in this region to be significantly higher than in the region we study in this paper. The disc composition is initially fully atomic and we assume a carbon abundance (by number) relative to hydrogen of $1.4 \times 10^{-4}$, an oxygen abundance of $3.2 \times 10^{-4}$, and a dust to gas ratio of 0.01. As the gas rotates in the potential, spiral structure develops and molecules form in the dense gas.

\begin{table}
	\centering
	\caption{Simulation parameters}
		\begin{tabular}{l c c }
   	         \hline
	         \hline
	          Simulation & Surface Density & Radiation Field Strength \\
	           & (\msun pc$^{-2}$) & (relative to Draine 1978) \\
	         \hline
	         Milky Way  & 10  & 1 \\
	         Low Density  & 4 & 1 \\
	         Strong Field & 10 & 10 \\
	         Low \& Weak & 4 & 0.1 \\
   	         \hline
	         \hline	         
		\end{tabular}
	\label{sims}
\end{table}

\begin{figure}
\begin{center}
\includegraphics[width=2.0in]{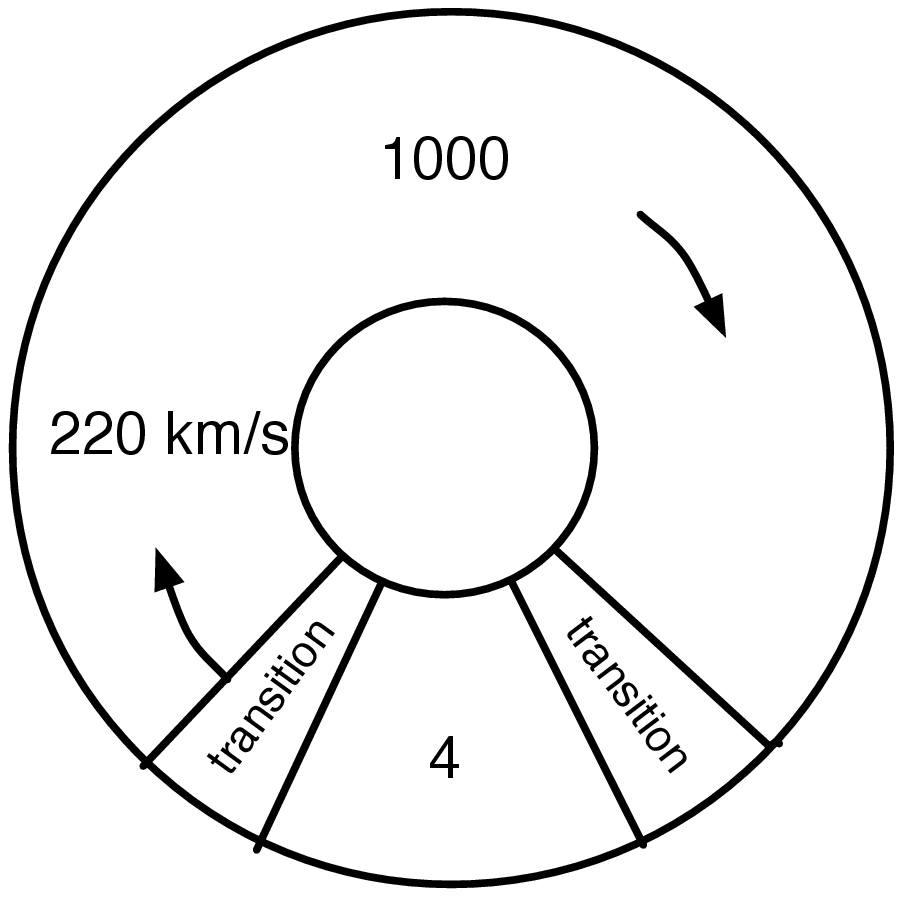}\\
\includegraphics[width=2.5in]{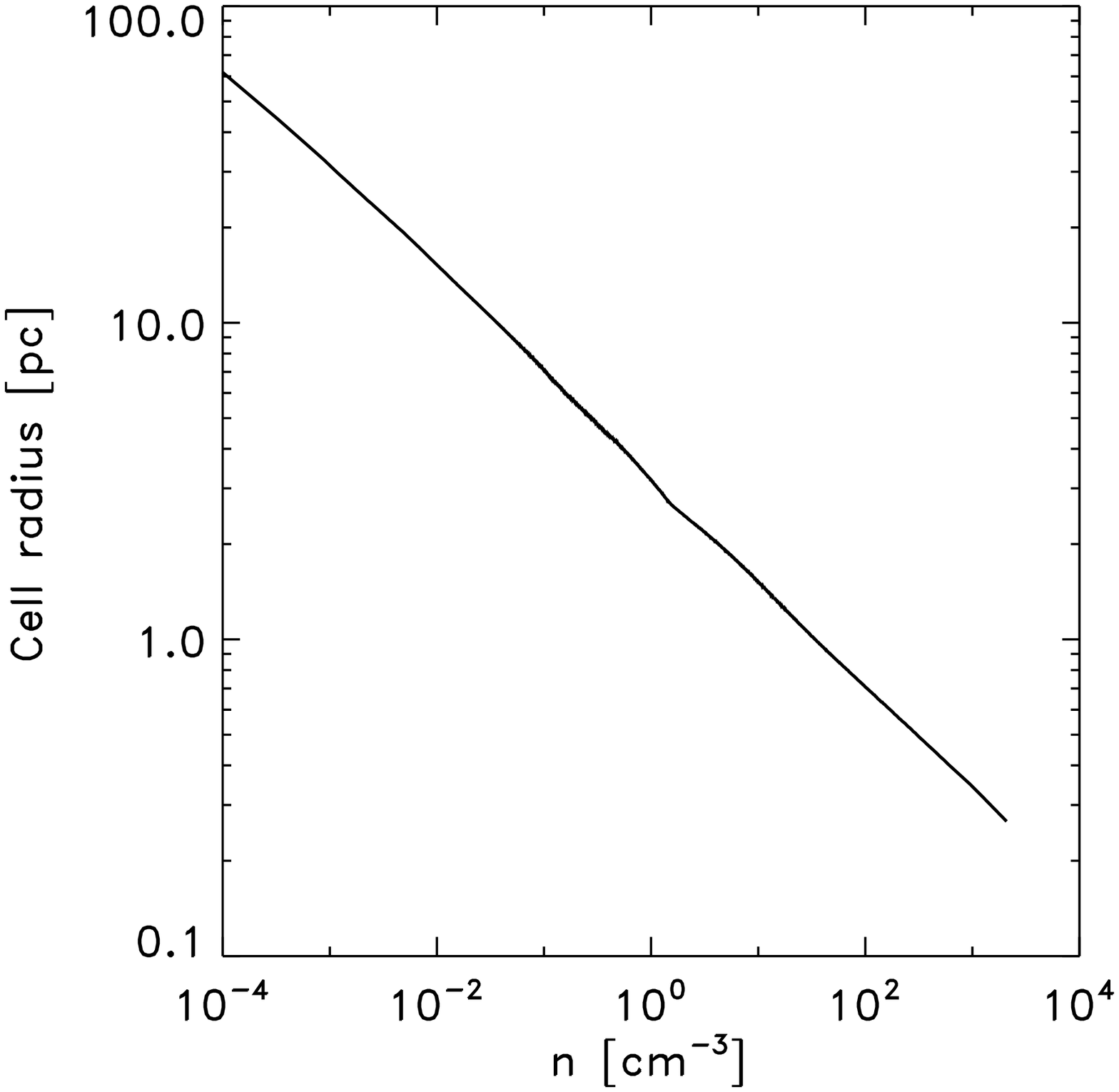}\\
\caption{\textit{Top}: Schematic diagram illustrating the refinement scheme. The numbers in each section represent the mean mass of gas in each \arepo cell. The sections all rotate  with the mean gas motion at 7.5 kpc. We focus our analysis on the highly refined 4 \msun section after the galaxy has completed one and a half full rotations. \textit{Bottom}: Spatial resolution as a function of gas number density. Gas with a number density of $10^3$ \cmc or more is resolved with a cell radius of better than 0.3 pc.}
\label{cartoon}
\end{center}
\end{figure}

We perform four simulations, with parameters as outlined in \tab \ref{sims}. The first of these is chosen to resemble the Milky Way. The initial gas surface density is 10 \sd and the strength of the UV field and the cosmic ray ionisation rate is set to reasonable Galactic values, $G_{0} = 1.7$ Habing units and $\zeta_{\rm H} = 3 \times 10^{-17} \: {\rm s}^{-1}$, respectively \citep{Draine78,vv2000}. The Low Density simulation has the same radiation field but a lower gas surface density, 4 \sd. This is chosen to be representative of a lower surface brightness galaxy. In the other two simulations, we explore the effects of varying the interstellar radiation field. The Low \& Weak simulation has a surface density of 4 \sd, but a UV field and cosmic ray ionisation rate that are an order of magnitude lower than in the fiducial Milky Way simulation. Finally, the Strong Field simulation has a surface density of 10 \sd and a UV field and cosmic ray ionisation rate that are an order of magnitude higher than the fiducial case.

In order to accurately model the formation of H$_2$ and CO, it is necessary to resolve individual molecular clouds in some detail. This is done using the system of mass refinement within \arepo. We select one section of the disc to have a cell mass of $4 \: {\rm M_{\odot}}$, but for efficiency reasons limit the cell mass to 1000\msun in the majority of the disc. Between the high resolution region and the lower resolution region we have buffer zones in which the resolution is smoothly increased from the lower value to the higher value. This avoids any discontinuous jumps in the effective resolution of the simulation. As the simulation evolves, the refined quadrant rotates along with the gas at a radial velocity of 220~\kms centred on a point at a radius of 7.5~kpc in the middle of the disc. The disc is then evolved for one and a half complete rotations, so that all of the gas has undergone six passages through the spiral features in the potential. This corresponds to a time $t = 261.1 \: {\rm Myr}$ after the beginning of the simulation. By this stage, the gas has developed dense spiral arms and complex filamentary structures. \fig \ref{cartoon} shows a cartoon of this refinement process. We concentrate our analysis on the most highly resolved section of the disc. \fig \ref{cartoon} also shows the spatial resolution as a function of number density in this region. Gas with a number density greater than $n = 1000 \: {\rm cm^{-3}}$, typical of gas in dense molecular clumps \citep{Bergin07}, is resolved at spatial scales of $\sim 0.3$ pc. A molecular cloud containing $10^4$ \msun would be resolved with 2500 cells in our model, which we will show in the next section is sufficient to properly recover the H$_2$ content of the cloud.

\section{Results from the Milky Way simulation}  \label{MW}

\subsection{Surface densities}

\begin{figure*}
\begin{center}
\includegraphics[width=6in]{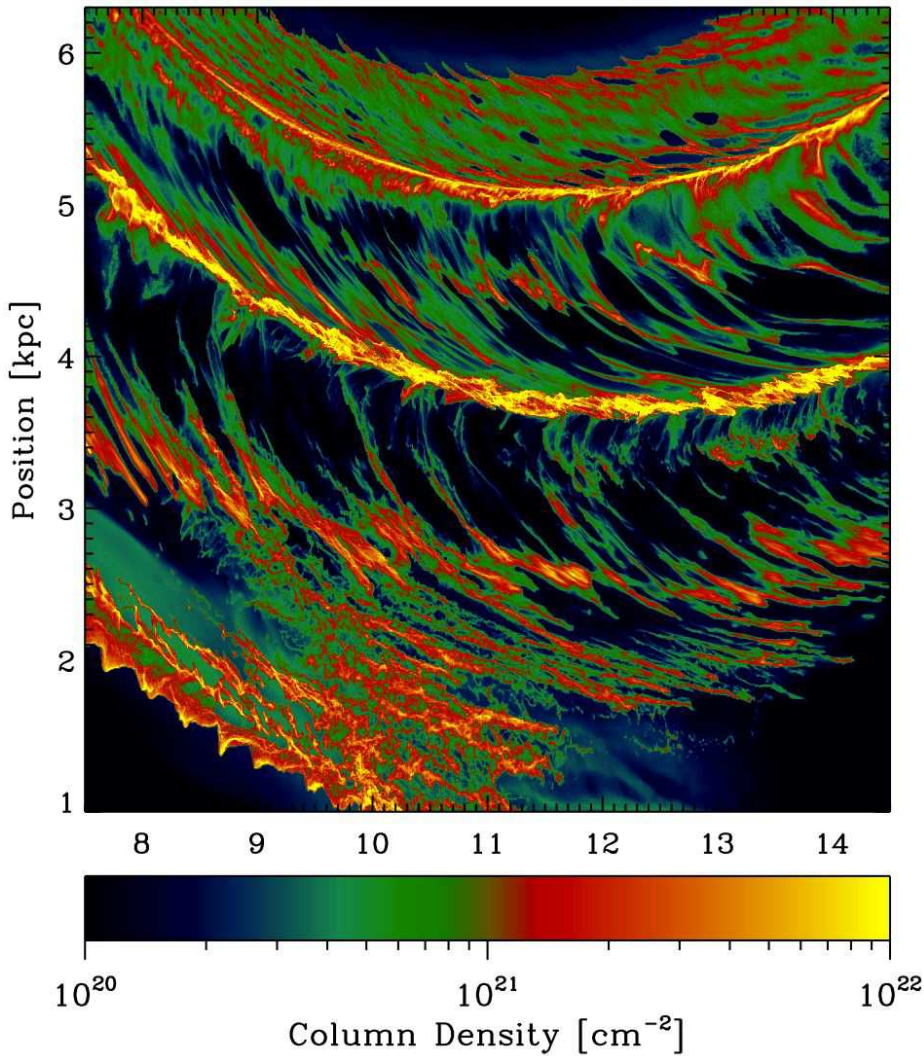}
\caption{Map of total column density of hydrogen nuclei for the highly resolved section of the disc in the Milky Way simulation. The gas has a range of morphologies, from dense spiral arms, to filamentary spurs, to diffuse inter-arm regions.}
\label{ColumnDensity}
\end{center}
\end{figure*}

\begin{figure}
\begin{center}
\includegraphics[width=3in]{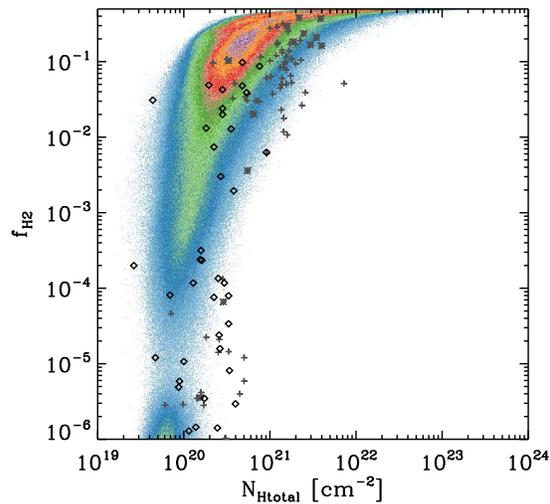}
\caption{Total column density of hydrogen nuclei versus the fractional H$_2$ abundance in the Milky Way simulation. The plotted points are a two dimensional histogram of the gas distribution with the blue points representing the least populated parts of the distribution and purple the most. The H$_2$ abundance as a function of hydrogen column density is consistent with the Galactic observations presented in \citet{Savage77} (\textit{grey crosses}), \citet{Wolfire08} (\textit{grey asterisks}) and \citet{Gillmon06} (\textit{black diamonds}), giving us confidence that we are resolving the clumping of gas in the simulation sufficiently well to accurately represent its chemical makeup.}
\label{fH2}
\end{center}
\end{figure}

In order to convert from the intrinsic 3D densities of the simulation into surface densities that can be compared to observations, we calculate the projected surface densities on a 2000 by 2000 pixel grid oriented parallel to the plane of the disc. The size of each individual pixel is therefore 3.5~pc by 3.5~pc. The column densities of H{\sc i}, H$_{2}$, CO etc.\ associated with each pixel can then be calculated simply by interpolating values from the \arepo mesh onto a series of rays oriented perpendicularly to the grid cells and then integrating along these rays. In regions of the disc dominated by cold, dense gas, the actual resolution of our hydrodynamical simulation can be much better than 3.5~pc, as Figure~\ref{cartoon} demonstrates. To ensure that any features missed by our gridding procedure do not affect the results, we have calculated the projected surface densities for two 700~pc by 700~pc sub-regions of the grid at ten times higher resolution (i.e.\ a pixel size of 0.35~pc by 0.35~pc). The surface density distributions derived from these fine grids are very similar to those found within the same regions of our coarse grid, suggesting that the results we present here are not particularly sensitive to our gridding procedure.

As an example of the results we obtain from our standard grid, we show in \fig \ref{ColumnDensity} a map of the total column density in the high-resolution section of the Milky Way simulation. We see from the map that the gas exhibits very different morphologies, ranging from dense spiral arms, to filamentary spurs, to diffuse inter-arm regions. Each of these regions has a different degree of shielding to the ambient radiation field and consequently a different molecular hydrogen abundance. 

\begin{figure*}
\begin{center}
\includegraphics[width=6in]{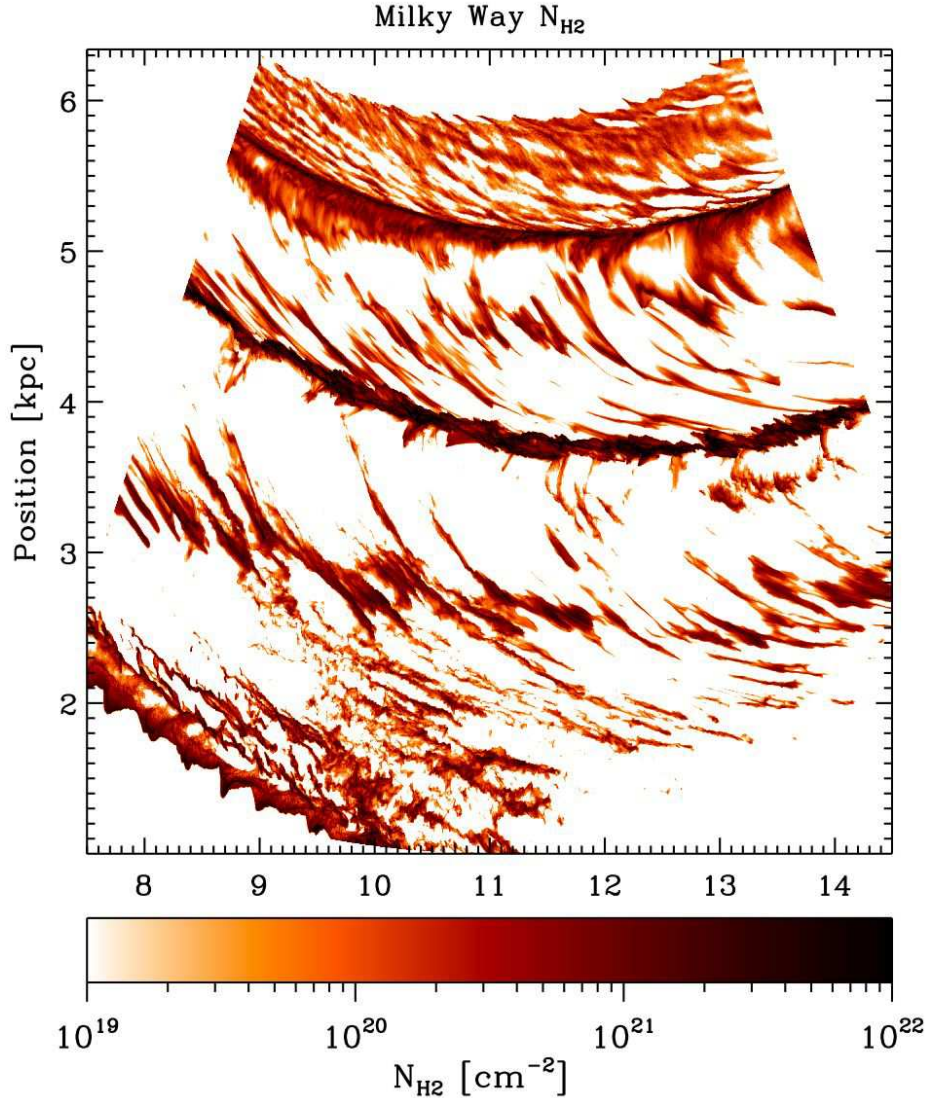}
\caption{Map of H$_{2}$ column density for the highly resolved section of the disc in the Milky Way simulation. H$_2$ is predominantly found in the spiral arms and in long filaments in the inter-arm regions.}
\label{maps}
\end{center}
\end{figure*}

\fig \ref{fH2} shows the fractional abundance of molecular hydrogen relative to hydrogen in all forms as a function of column density. In this work, we define the fractional abundance of H$_{2}$ via the relationship $f_{\rm H_{2}} \equiv n_{\rm H_{2}} / n$, where $n_{\rm H_{2}}$ is the number density of hydrogen molecules and $n \equiv 2 n_{\rm H_{2}} + n_{\rm H} + n_{\rm H^{+}}$ is the total number density of hydrogen nuclei. With this definition, the maximum value of the fractional abundance is $f_{\rm H_{2}} = 0.5$, corresponding to fully molecular hydrogen. Between column densities of $10^{20}\, $\cms and $10^{21} \,$\cms the molecular hydrogen begins to self-shield and its abundance rises dramatically. A similar jump in molecular hydrogen abundance is seen observationally at similar total column densities, as shown by \citet{Leroy07} and \citet{Wolfire08}. 

\citet{Gnedin09} presented a galactic scale model of molecular hydrogen formation in which these observations were used to calibrate a clumping factor, used to account for small-scale, unresolved density fluctuations, and tuned to ensure that the model matched observations. Our results in \fig \ref{fH2} are a good match to the observed transition without us having to apply any calibration factors. There is some suggestion in \fig \ref{fH2} that our column densities are slightly lower for a given value of $f_{\rm H_{2}}$ than some of the observational data \citep[e.g.][]{Savage77}. However, these observations were taken along long sight-lines within the galactic disc which will have higher column densities than in our face-on disc. The observations of \citet{Gillmon06} along sight-lines perpendicular to the disc (shown by the bold diamonds in \fig \ref{fH2}) are in good agreement with our data. This gives us confidence that the small-scale galactic structure is sufficiently resolved to accurately describe its chemical makeup.

\begin{figure}
\begin{center}
\includegraphics[width=3.5in]{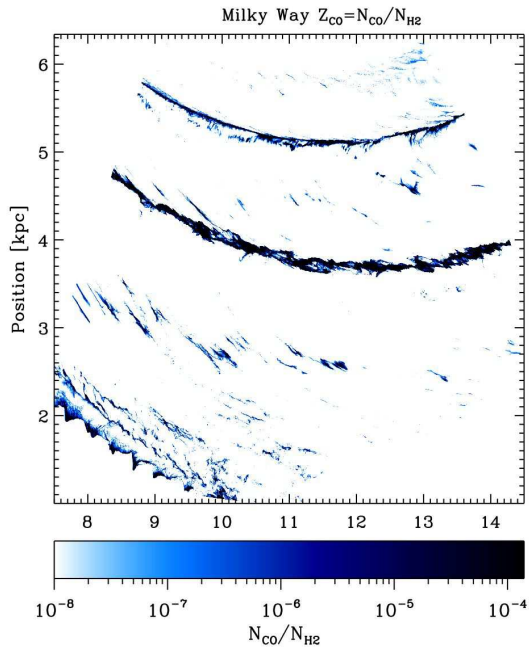}
\caption{Map of the CO/H$_2$ column density ratio in the highly resolved segment of the Milky Way simulation. CO is predominantly found in the spiral arms and does not trace the more diffuse H$_2$ in the inter-arm regions well.}
\label{h2-co}
\end{center}
\end{figure}

\fig \ref{maps} shows the column density of molecular hydrogen in the highly resolved disc segment. Molecular hydrogen is predominantly present in the spiral arms, but there is also molecular gas in inter-arm spurs and in the inner regions of the disc. In the inter-arm regions molecular hydrogen is often found in long filaments that were originally spurs connected to the spiral arms but that were sheared off as the disc rotated. \fig \ref{h2-co} shows the ratio of H$_2$ to CO column densities in the gas. There is considerable variation in the abundance of CO. In particular, the long inter-arm filaments, which are so apparent in \fig \ref{maps}, are much less visible in CO. This can be attributed to their narrow filamentary geometry being inefficient at shielding the gas from the ambient radiation field. Due to the low abundance of CO in these regions, the molecular gas there is likely to appear `dark' in observations of CO emission.

\subsection{The relationship between CO and H$_{2}$ column densities}
Although our simulations provide us with information on the full 3D distributions of the H$_{2}$ and CO abundances, in general these are not observable quantities. For comparison with observations, it is more useful to examine the correlation between the H$_{2}$ and CO column densities, and the column-averaged abundance of CO relative to H$_{2}$, $Z_{\rm CO} = N_{\rm CO} / N_{\rm H_{2}}$.

\begin{figure*}
\begin{center}
\begin{tabular}{c c}
\includegraphics[width=2.7in]{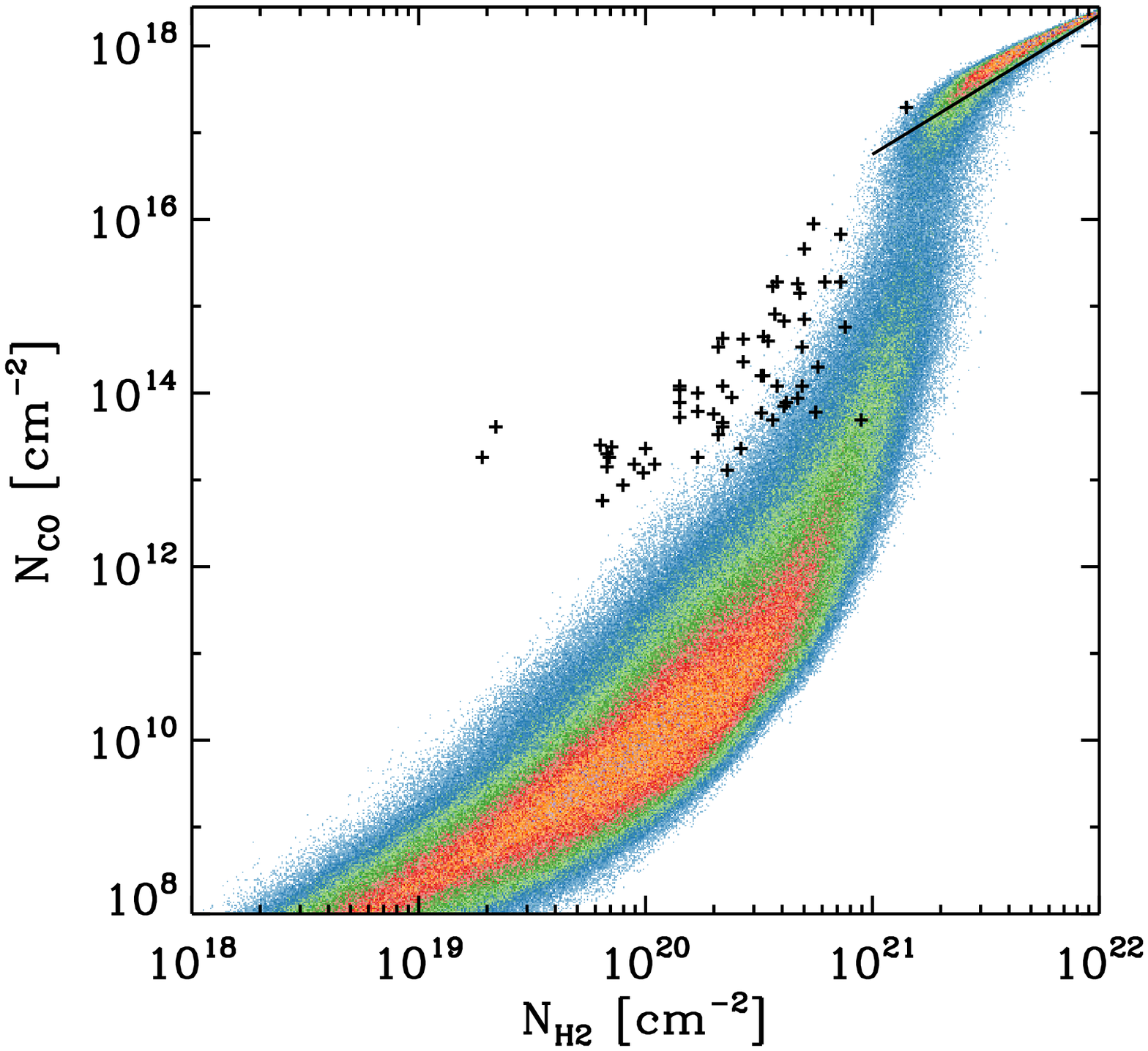}
\includegraphics[width=2.7in]{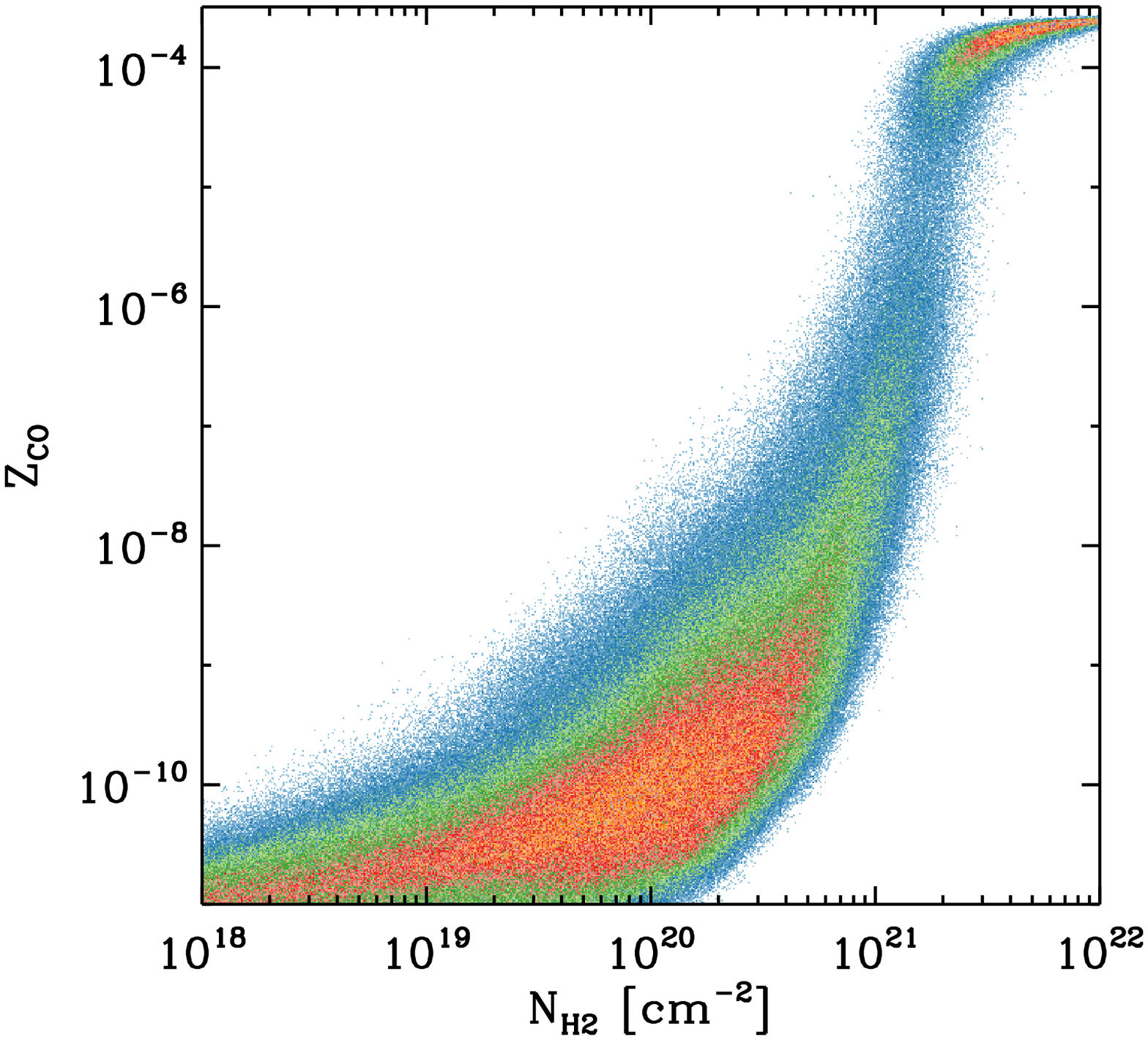}\\
\end{tabular}
\caption{Relationship between \NHm, \NCO and \ZCO in the Milky Way simulation. The plotted points are a two dimensional histogram of the gas distribution, with the blue points representing the least populated parts of the distribution and orange the most densely populated. The crosses show observations by \citet{Sheffer08} of \NCO along diffuse sight lines, and the solid line shows a fit to a synthesis of \NCO observations in dark clouds by \citet{Federman90}. At low column densities there is considerable local variation in the amount of CO present in the gas, as shown by the large scatter in the distributions at these column densities.}
\label{sheffer}
\end{center}
\end{figure*}

The left panel of \fig \ref{sheffer} shows the relation between \NHm and \NCO in our fiducial Milky Way simulation. At high H$_{2}$ column densities ($N_{\rm H_{2}} > 10^{21} \: {\rm cm^{-2}}$), our simulations produce slightly more CO than predicted by \citet{Sheffer08} from their measurements of UV absorption. In part, this is due to our use of the \citet{Nelson97} chemical network, which tends to overproduce CO in dense clouds \citep{Glover12a}, but it should also be noted that the \citet{Sheffer08} observations also include some upper limits. At lower column densities, we see that an increasingly large discrepancy develops between the observed and simulated CO column densities: there seems to be considerably more CO present in the actual diffuse ISM than is produced in our models. This discrepancy is not unique to our model -- other authors have previously discussed the difficulties involved in reproducing the observed \NCO--\NHm relationship at low $N_{\rm H_{2}}$ with standard PDR models \citep[see e.g.\ the detailed discussion in][]{Sonnentrucker07}, a problem which may point to the importance of non-thermal chemistry in these regions \citep{Zsargo03,Sheffer08}. Fortunately, we see large differences between our results and the \citet{Sheffer08} observations only for CO column densities $N_{\rm CO} < 10^{15} \: {\rm cm^{-2}}$. At column densities typical of molecular clouds we reproduce well the findings of an synthesis of dark cloud observations by \citet{Federman90}. Since gas at low column densities will not produce significant CO emission, the fact that we underestimate the amount of CO present at these columns should not strongly affect the results we derive later for the dark gas fraction.

In the right panel of \fig \ref{sheffer}, we show how $Z_{\rm CO}$ behaves as a function of the H$_{2}$ column density. At low column densities, the abundance is much lower than the value of $\sim 10^{-4}$ commonly assumed for GMCs. As can be seen from the scatter in the distribution of \ZCO at H$_2$ column densities of around $10^{20}$\cms, there is considerable variability in this column density regime. Similar regional variability in CO abundances has been seen in diffuse gas clouds by \citet{Liszt12}. 

\subsection{CO emission and ${X_{\rm CO}}$} \label{xfactor}

Generating full synthetic images of the CO emission produced by the gas in our simulations with a resolution that is well matched to our hydrodynamical resolution is a very challenging computational problem that lies outside the scope of this initial study. Nevertheless, in order to properly identify CO-dark gas, we clearly need to have some idea of where the bulk of the CO emission is produced. We have therefore chosen to make use of a simplified curve-of-growth approach \citep[see e.g.][]{Pineda08,Glover11}. This is not full radiative transfer, but it does allow us to estimate the velocity-integrated intensity in the $J = 1 \rightarrow 0$ line of CO. 

We begin by estimating the mean temperature of the CO-emitting gas in each pixel of our grid by computing a simple CO abundance-weighted average of the temperatures of the \arepo cells contributing to that grid point:
\begin{equation}
T_{\rm CO} = \frac{\sum \limits_{i=0}^n T_i A_{{\rm CO}, i}}{\sum \limits_{i=0}^n A_{{\rm CO}, i}},
\end{equation}
where $A_{{\rm CO}, i}$ and $T_{i}$ are the fractional abundance of CO and the temperature in cell $i$, respectively. Note that as we are using a refinement scheme that attempts to keep the mass in each \arepo cell constant, this abundance-weighted average is essentially equivalent to an average weighted by the mass of CO in each cell. 

Next, we need to relate the excitation temperature of the CO to $T_{\rm CO}$. For simplicity, we assume that the CO is in local thermodynamic equilibrium (LTE), with an excitation temperature $T_{\rm ex} = T_{\rm CO}$. This is a reasonable assumption for our high column density sight-lines, as these primarily probe regions with high gas densities and high optical depths in the CO lines. Along more diffuse sight-lines, however, our assumption leads to us overestimating the excitation temperature. For example, \citet{Burgh07} found an average excitation temperature of $T_{\rm ex} = 4.1$~K along diffuse CO sight-lines in the Milky Way. This is significantly lower than the gas temperature, which in our simulations has a typical CO-weighted value of $T_{\rm CO}=25$~K for CO column densities in the range $10^{15} < N_{\rm CO} \le 10^{16}$\cms. Such an assumption would have the effect of increasing the emission and would therefore make our derived dark gas fraction a conservative estimate. However, as we will see later, our results are not particularly sensitive to this choice.

The next step in producing our synthetic images involves estimating the optical depth in the CO $J = 1 \rightarrow 0$ line. This is related to the CO column density via
\begin{equation}
\tau_{10} = \frac{A_{10}c^3}{8\pi \nu_{10}^3} \frac{g_1}{g_0} f_0 \left [ 1- \exp \left ( \frac{-E_{10}}{kT_{\rm ex}} \right)  \right ] \frac{N_{\rm CO}}{\Delta v}
\end{equation}
where $A_{10}$ is the spontaneous radiative transition rate for the $J = 1 \rightarrow 0$ transition, $\nu_{10}$ is the frequency of the transition, $E_{10} = h\nu_{10}$ is the corresponding energy, $g_0$ and $g_1$ are the statistical weights of the $J = 0$ and $J = 1$ levels, respectively, and $f_0$ is the fractional level population of the $J = 0$ level. We take values for $A_{10}$ and $\nu_{10}$ from the Leiden Atomic and Molecular Database (LAMDA) \citep{Schoier05}. Our assumption that the CO molecules are in LTE means that $f_{0}$ is given by the simple expression $f_{0} = 1 / Z(T)$, where $Z$ is the partition function of the CO molecule. Finally, following the arguments presented in \citet{Bolatto13} we obtain an estimate of the velocity dispersion of the gas from the virial theorem. Observations of CO-bright molecular clouds show that they frequently have velocity dispersions close to those that they should have in virial equilibrium, although one cannot necessarily conclude from this that the clouds actually are in equilibrium \citep[see e.g.][]{Ballesteros06}. Our simulated clouds clearly are not in virial equilibrium, since our simulations do not include the effects of self-gravity. Nevertheless, a virial estimate should still give us a reasonable guide to the expected velocity dispersion of the gas. We therefore assume that $\Delta v= \sqrt{GM/5 r_{\rm pix}}$, where $r_{\rm pix}$ is the radius of the \arepo pixel. For the dense, highly molecular gas this results in typical values of around 3 \kms, which are similar to the values used in \citet{Glover10} for dense clouds. As we will see shortly, varying this number by a factor of a few would not significantly change our results. For a temperature of $T=10$~K, typical of cold molecular gas, and for a velocity of 3 \kms, we find that  the CO becomes optically thick at a CO column density $N_{\rm CO} \sim 2 \E^{16} $\cms.

Finally, once we have $\tau_{10}$, we compute $W_{\rm CO}$ using the curve-of-growth method of \citet{Pineda08}, which relates the integral of the photon escape probability $\beta(\tau)$ to the integrated CO intensity:
\begin{equation}
W_{\rm CO} = 2 T_{\rm CO} \Delta v \int_0^{\tau_{10}} \beta(\tau) d\tau.
\end{equation}
We approximate $\beta(\tau)$ by assuming that it is the same as for a plane-parallel, uniform slab \citep{Tielens05}:
\[
\beta(\tau) =
 \begin{cases}
  [1 - \exp(-2.34 \tau)] / 4.68 \tau & \text{if } \tau \le 7 \\ 
  ( 4 \tau [ \ln (\tau/\sqrt{\pi})]^{1/2})^{-1} & \text{if } \tau > 7
  \end{cases}
  \]
We integrate the escape probability for each grid cell from zero to its optical depth and derive the estimated value of \WCO associated with each value of \NCO and \NHm. 

\begin{figure*}
\begin{center}
\includegraphics[width=6in]{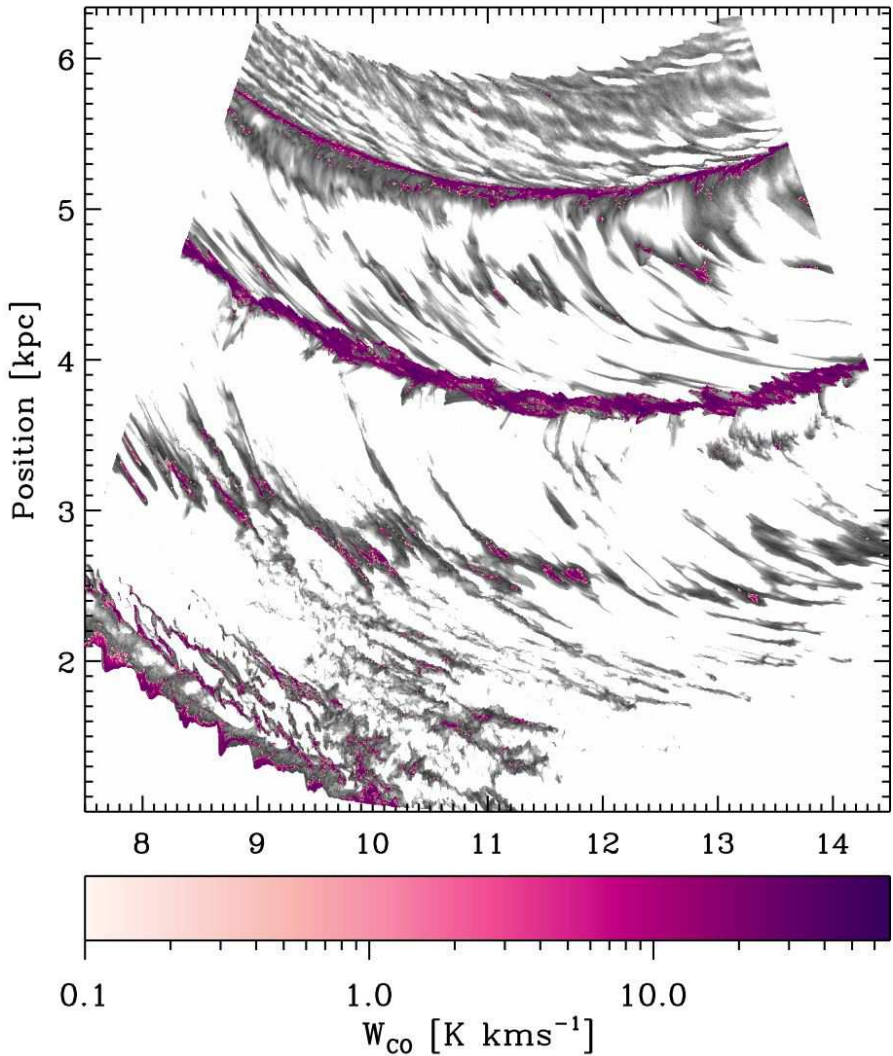}
\caption{Morphology of the molecular gas in our Milky Way simulation. The greyscale background image shows the H$_{2}$ column density (c.f.\ Figure~\ref{maps}), while the 
purple points show the strength of the CO velocity-integrated intensity, \WCO, estimated as described in the text. Many of the clouds in the inter-arm region have no portions with integrated intensities above 0.1~K~km~s$^{-1}$ and thus would appear entirely `dark' in CO observations.}
\label{morphology}
\end{center}
\end{figure*}

In Figure~\ref{morphology}, we present the final synthetic image of \WCO, overlaid on a map of the H$_{2}$ column density. We show only the emission from regions with integrated intensities $W_{\rm CO} \ge 0.1 \: {\rm K \: km \: s^{-1}}$, corresponding approximately to a typical detection sensitivity for large-scale CO mapping. We see that the spatial distribution of the CO emission is well-correlated with the H$_{2}$ in the dense spiral arms. However, in the inter-arm regions there are large clouds of H$_{2}$ that are not well traced by CO. We will return to this point later, in Section~\ref{morph}.

\begin{figure}
\begin{center}
\includegraphics[width=3in]{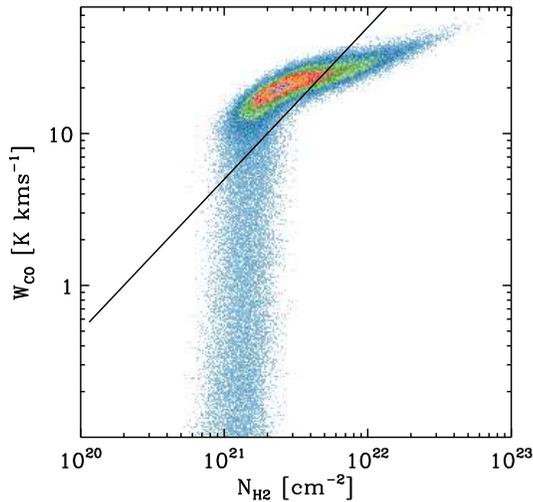}
\caption{Integrated intensity in the $J = 1 \rightarrow 0$ line of CO, \WCO, as a function of \NHm in our Milky Way simulation. The points are a 2D histogram of the gas distribution with blue representing the least populated part of the distribution and orange the most. The solid line shows what would be expected if the value on small scales was simply the canonical Galactic value of $X_{\rm CO}= 2.0\E^{20}$ cm$^{-2}$K$^{-1}$km$^{-1}$s. The actual mean value is only slightly larger than this, $X_{\rm CO}= 2.2\E^{20}$ cm$^{-2}$K$^{-1}$km$^{-1}$s, but there is clearly significant small-scale variation in the value of \XCO.}
\label{mw_xfactor}
\end{center}
\end{figure}

Using our map of the integrated intensity, it is also possible to calculate the CO-to-H$_2$ conversion factor \XCO=\NHm/\WCO for each cell. \fig \ref{mw_xfactor} shows \WCO as a function of \NHm, with the solid line showing what we would obtain if we simply adopted the standard Galactic value of $X_{\rm CO, gal}= 2\E^{20}$ cm$^{-2}$K$^{-1}$km$^{-1}$s in each pixel \citep{Bolatto13}. We see that on a pixel-by-pixel basis, the distribution of integrated intensities differs greatly from the simple estimate $N_{\rm H_{2}} / X_{\rm CO, gal}$. At low H$_{2}$ column densities, there is much less emission than this estimate predicts, owing to the influence of CO photodissociation, while at high H$_{2}$ column densities, the emission is suppressed by the high CO line opacity. Similar results have previously been found for small-scale cloud models \citep[see e.g.][]{Shetty11b} and for observations of individual GMCs \citep[see e.g.][]{Pineda08,Lee2014}, underscoring the fact that $X_{\rm CO}$ only becomes a meaningful quantity when one averages the emission and the H$_{2}$ column density on scales comparable to or larger than the typical size scale of individual GMCs \citep[see also the discussion of this point in][]{Bolatto13}. Indeed, if we average the emission and the H$_{2}$ column density over our high-resolution region everywhere \WCO is potentially observable (i.e.\ wherever $W_{\rm CO} > 0.1 \, {\rm K \, km \, s^{-1}}$), we obtain a value $X_{\rm CO} = 2.2 \times 10^{20}$~cm$^{-2}$K$^{-1}$km$^{-1}$s, very close to the fiducial Milky Way value. This estimate of $X_{\rm CO}$ is probably the most appropriate to compare with observational determinations of the same value, as these are typically made using observations of CO-bright clouds, where both the H$_{2}$ column density and the CO integrated intensity can be determined. If we instead average over all of the gas in our high-resolution region, including the CO-dark clouds, we obtain a larger value, $X_{\rm CO} = 3.9 \times 10^{20}$~cm$^{-2}$K$^{-1}$km$^{-1}$s.


\subsection{Quantifying the dark gas fraction}
To quantify the amount of CO-dark molecular gas in our simulations, we define a dark gas fraction
\begin{equation}
f_{\rm DG}(x)=\frac{M^x_{\rm H_2}}{M^{\rm CO}_{\rm H_2}+M^x_{\rm H_2}},
\end{equation} 
where $M^x_{\rm H_2}$ is the mass of CO-dark H$_2$ with emission below an intensity of $x=W_{\rm CO}$, and $M^{\rm CO}_{\rm H_2}$ is the mass of CO-bright H$_2$ above this threshold. Our dark gas fraction therefore is the ratio of the CO-dark molecular gas to the total molecular gas. This definition is equivalent to that used in \citet{Wolfire10}, but differs from that used by some other authors, who define the dark gas fraction relative to the total gas mass (i.e.\ the sum of the atomic and molecular masses).

\begin{figure}
\begin{center}
\includegraphics[width=3in]{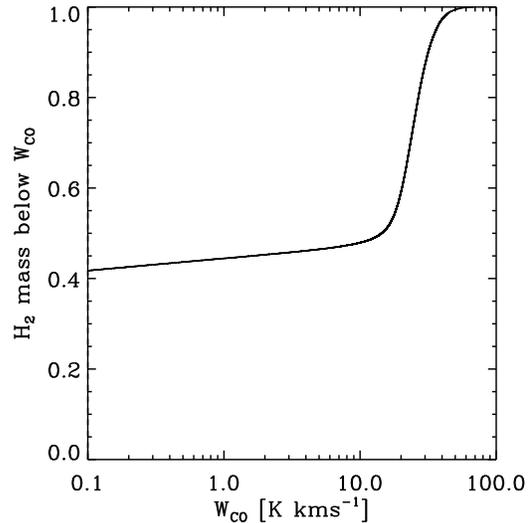}
\caption{Dark gas fraction as a function of \WCO. Approximately 42\% of the total molecular gas mass is found in regions with $W_{\rm CO} \le 0.1$~K~\kms.}
\label{darkWCO}
\end{center}
\end{figure}

Clearly, in order to compute $f_{\rm DG}$, we need to be able to distinguish between CO-dark and CO-bright gas. However, it is not immediately obvious how to do this: how faint does the CO emission from a cloud of gas need to be before we call that gas CO-dark? In our analysis, rather than adopting an arbitrary emission threshold for distinguishing between CO-dark and CO-bright gas, we have instead computed $f_{\rm DG}$ as a function of the choice of threshold $(x)$. The results are plotted in Figure~\ref{darkWCO}. We see from the Figure that in practice, the value we derive for $f_{\rm DG}$ is not particularly sensitive to our choice of threshold, provided that we take a threshold $W_{\rm CO, th} < 10 \: {\rm K \: km \: s^{-1}}$. We find that 42\% of the total molecular mass is found in regions with very little CO emission (integrated intensities $W_{\rm CO} < 0.1\: {\rm K \: km \: s^{-1}}$) and thus the dark gas fraction in the disc is $f_{\rm DG} = 0.42$.\footnote{Proving once again that the answer to everything is 42 \citep{Adams79}.} 

Around 55\% of the total molecular gas mass in the simulation is found in bright regions with high CO integrated intensities, $W_{\rm CO} > 10 \: {\rm K \: km \: s^{-1}}$, corresponding to giant molecular clouds. Only a few percent of the total molecular gas mass is found in regions with integrated intensities in between these two limits, and so our choice of threshold has little influence on our results, provided that it lies somewhere in this range. In addition, this behaviour means that our derived dark gas fractions have little sensitivity to errors of a factor of a few in \WCO, and so even though our determination of the integrated intensities is highly approximate, our results for $f_{\rm DG}$ should be robust.

\begin{figure}
\begin{center}
\includegraphics[width=3in]{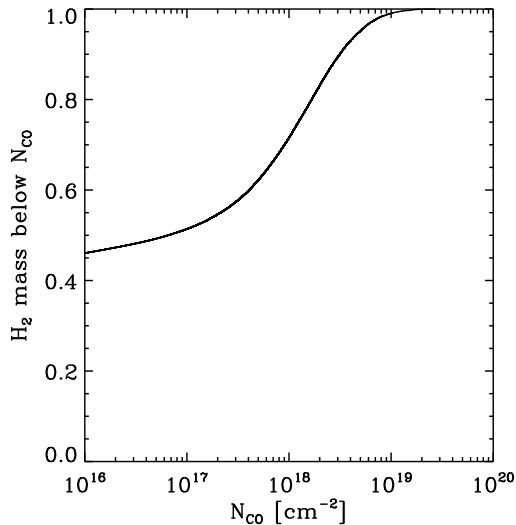}
\caption{Dark gas fraction as a function of the CO column density, \NCO. Half of the molecular gas is found in regions with $N_{\rm CO} \le 6.5 \times 10^{16} \: {\rm cm^{-2}}$.}
\label{darkgas}
\end{center}
\end{figure}

We have also examined how the H$_{2}$ in our simulations is distributed as a function of CO column density (Figure~\ref{darkgas}). Figure \ref{darkgas} shows that there are two distinct regimes, an initial smooth increase up to CO column densities of a few times $10^{17}$\cms, and then a rapid increase once we reach the $A_{\rm V} > 1$ regime where CO becomes abundant. 

\subsection{The morphology of the dark gas}\label{morph}

As well as allowing us to quantify the amount of CO-dark molecular gas we expect to find in the solar neighbourhood, our simulations also allow us to examine how this gas is distributed spatially. One obvious hiding place of this gas is in dark envelopes surrounding CO-bright molecular clouds. \citet{Wolfire10} used a spherical PDR model to quantify the amount of molecular gas that could be located in these envelopes and found that it could account for as much as 30\% of the mass of a typical GMC. Our simulations also find CO-dark H$_2$ in the immediate vicinity of CO-bright clouds, but in addition, they show that a significant amount of CO-dark H$_{2}$ is located in extended filaments with lengths of tens to hundreds of parsecs, located in the inter-arm regions of our simulated disc and inclined at a roughly 45\deg angle to the main spiral arms (Figure~\ref{morphology}). These filaments are created when denser gas clumps are sheared out by the differential rotation of the disc  following their passage through the spiral arm \citep[see e.g.][]{Dobbs06a,Bonnell13}. 

\begin{figure*}
\begin{center}
\begin{tabular}{c c}
\includegraphics[width=3.5in]{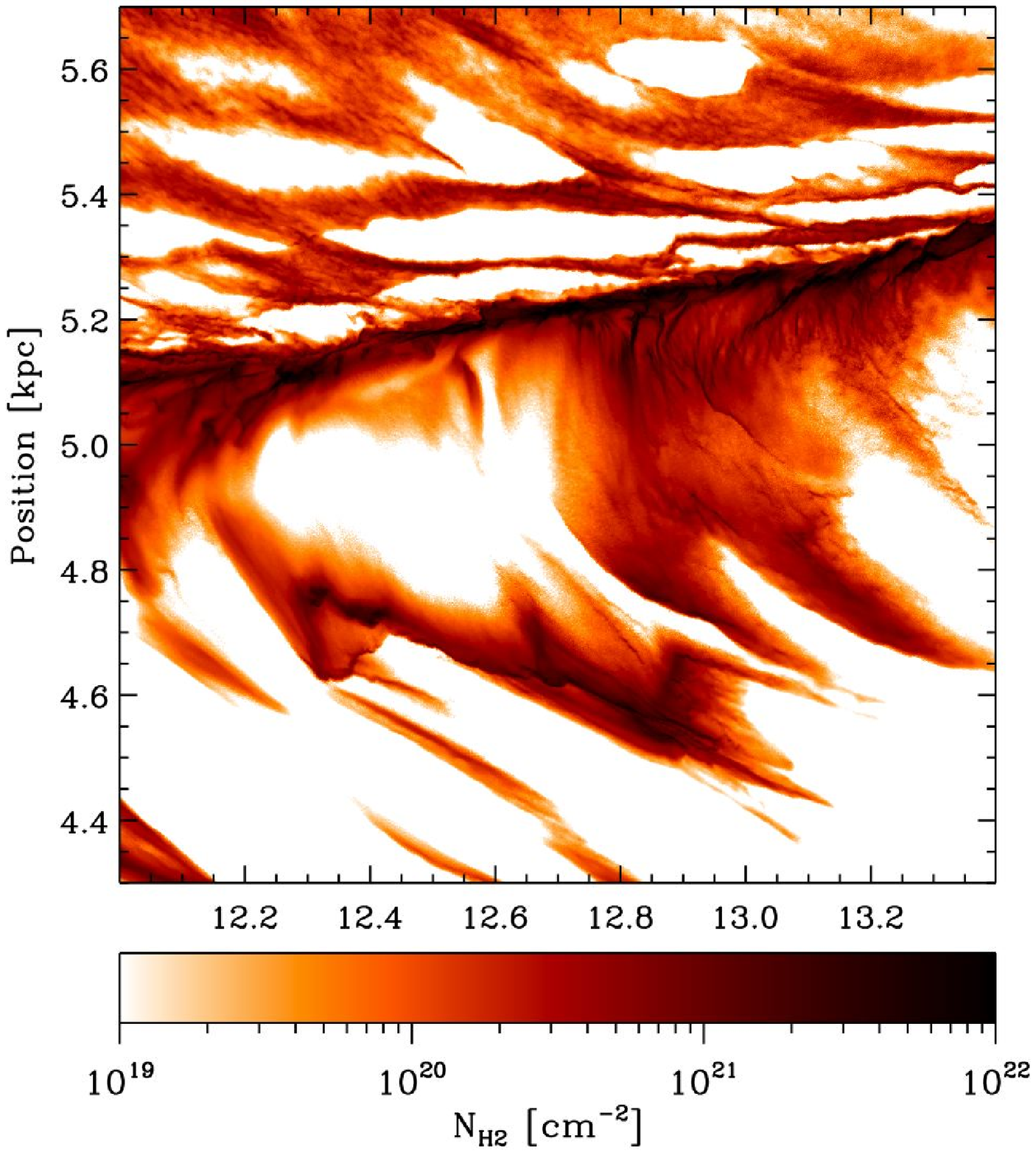}
\includegraphics[width=3.5in]{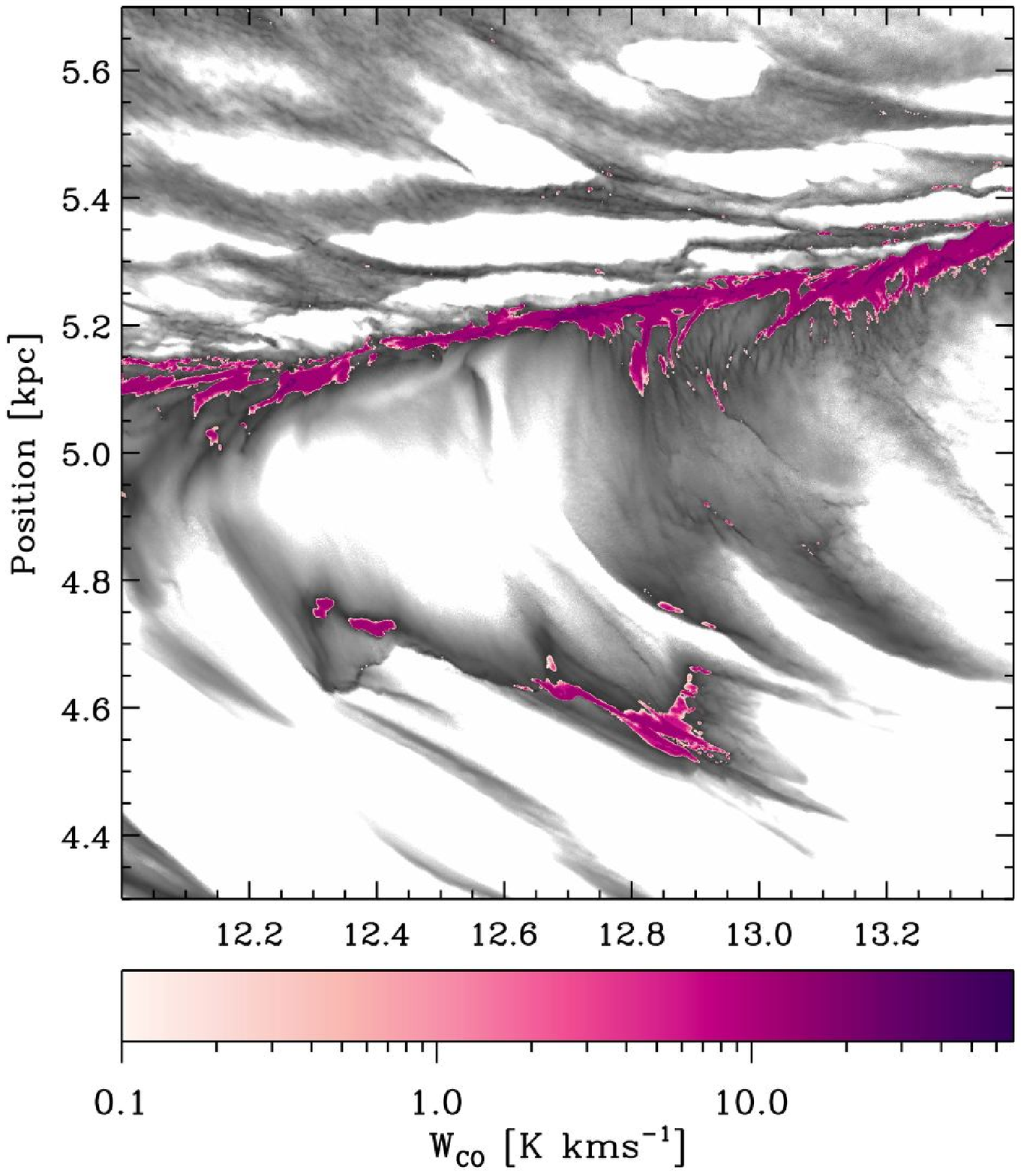}
\end{tabular}
\caption{Close-up view of the H$_2$ column density and CO integrated intensity produced in a spiral arm and an associated spur.}
\label{panel2}
\end{center}
\end{figure*}

\begin{figure*}
\begin{center}
\begin{tabular}{c c}
\includegraphics[width=3.5in]{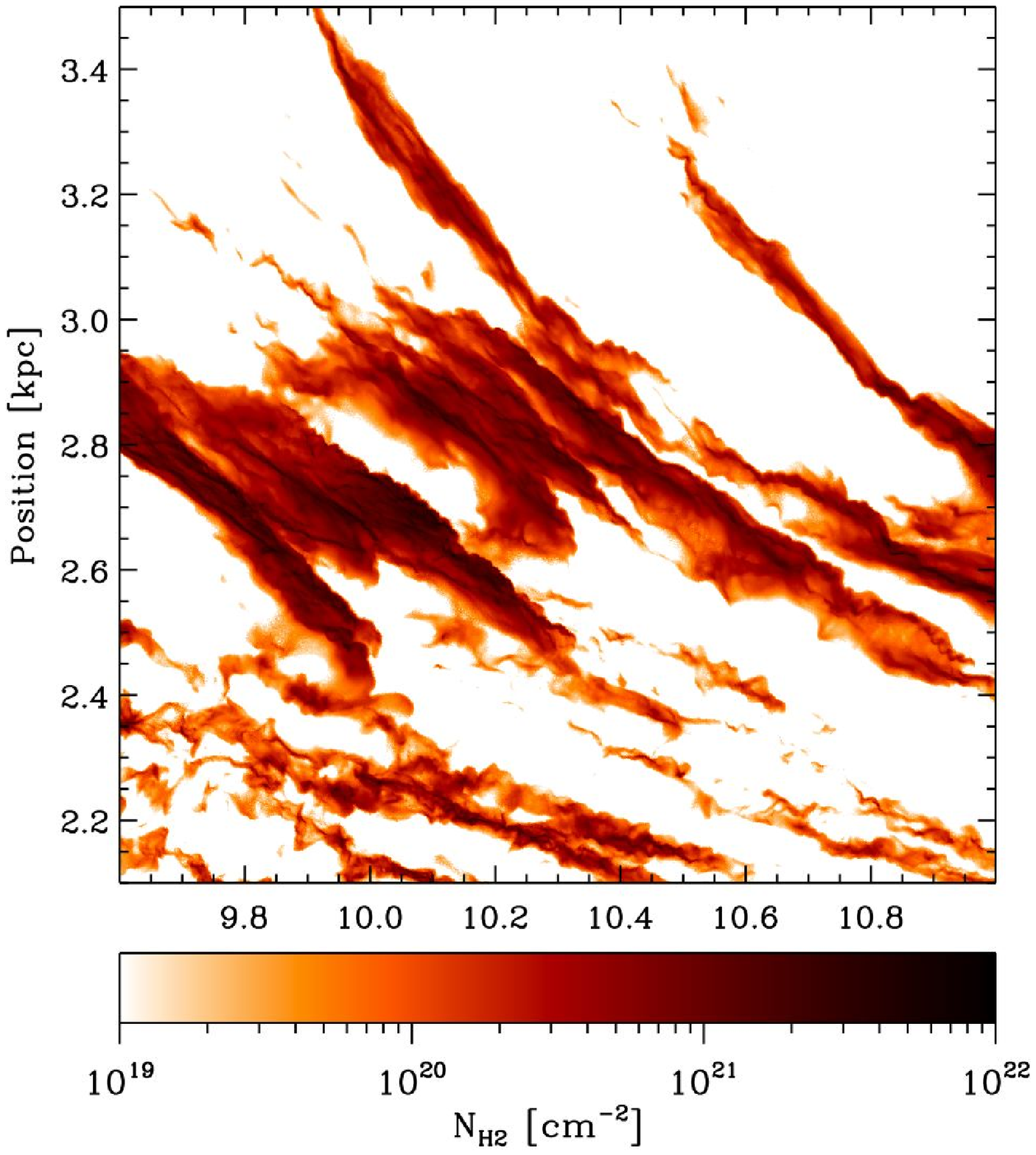}
\includegraphics[width=3.5in]{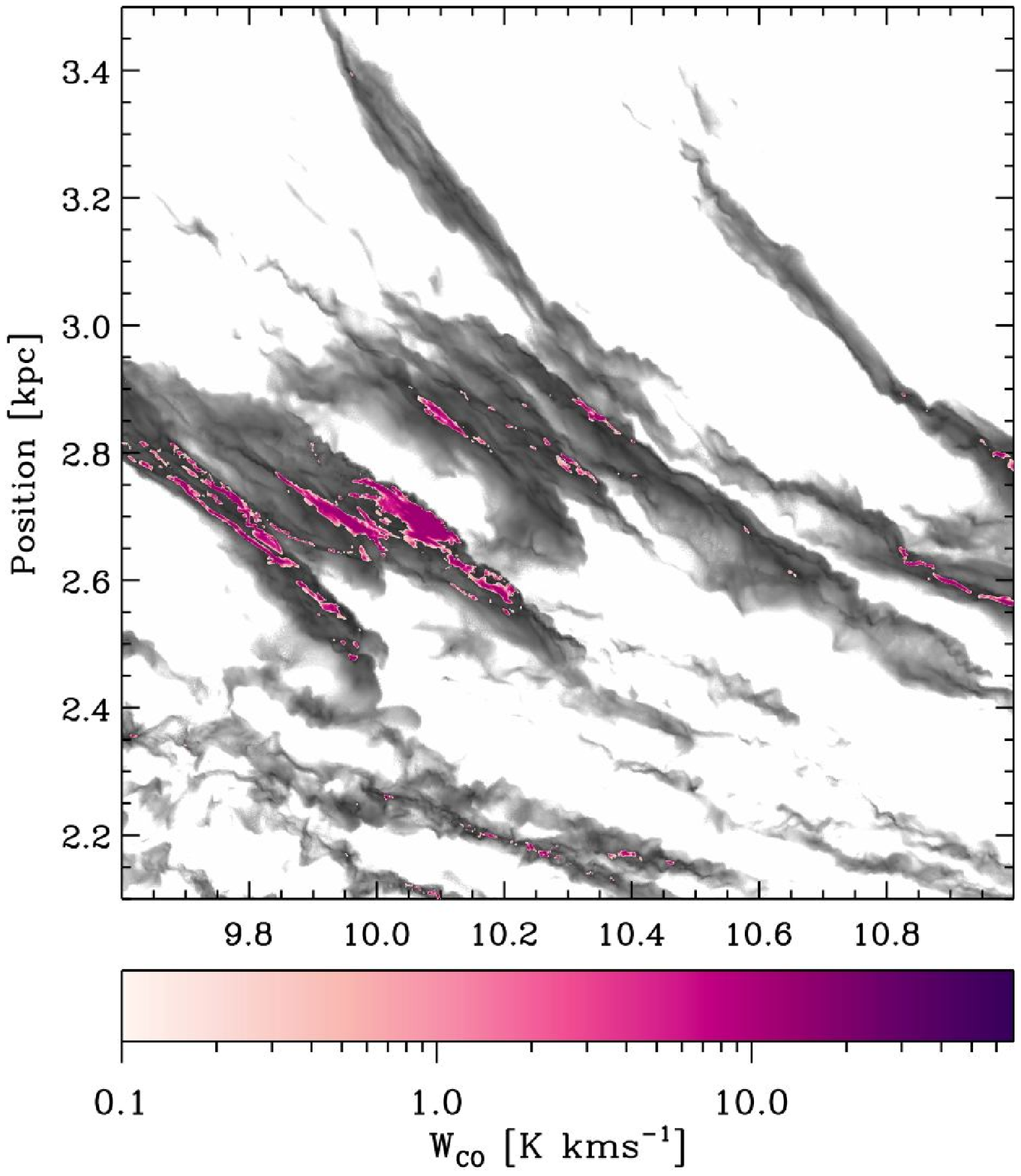}
\end{tabular}
\caption{Close-up view of the H$_2$ column density and CO integrated intensity from a region containing inter-arm filaments.}
\label{panel1}
\end{center}
\end{figure*}

The CO-dark regions have characteristic H$_{2}$ column densities in the range $10^{20}\: {\rm cm^{-2}} < N_{\rm H_{2}} < 10^{21} \: {\rm cm^{-2}}$, too low to allow them to retain much CO (c.f.\ Figure~\ref{sheffer}). Figure \ref{panel2} shows a close up view of the CO-bright and dark regions in a spiral arm and spur, and Figure \ref{panel1} shows the same for an inter-arm filament. While in spherical PDR codes the CO-dark gas naturally forms a uniform `skin' surrounding the cloud, Figures \ref{panel2} and \ref{panel1} show that the CO-dark cloud envelopes do not necessarily trace the outline of the CO-bright clouds that they surround. Particularly in the inter-arm regions, the dark gas is itself filamentary and can extend far beyond the boundaries of any CO-bright cloud which it envelopes. 

The geometry of the gas shown in \fig \ref{panel1} makes it more susceptible to dissociation since it will be easier for radiation to penetrate the cloud along the short axes of the filament. Compared to a spherical cloud of the same mass and mean density a filamentary cloud has a greater surface to volume ratio which increases the difficulty of achieving a sufficient column to shield the CO. Many of the longer filaments do have CO-bright regions close to their centres, but some remain CO-dark throughout. Absorption line measurements probing the ends of these structures would probably characterise them as diffuse molecular clouds \citep{Snow06}, while their central regions would be visible in CO emission and would be classified as dense molecular clouds, even though in reality both regions are part of the same physical structure. 

It has become increasingly obvious over the last few years that long filamentary molecular clouds are a common feature in the Milky Way \citep[e.g.][]{Henning10,Tackenberg13,Contreras13,Ragan14}. A good example of such a cloud is the ``Nessie'' filament \citep{Jackson10}, which has a length of at least 80 pc and is thought to coincide with spiral structure. Recent observations by \citet{Li13} have found an even larger $\sim 500$ pc filamentary `wisp' of molecular gas that has a coherent velocity structure. Observations of external disc galaxies find similar features. For example the PAWS survey of CO emission in M51 \citep{Schinnerer13} shows long filaments and spurs coming out of the spiral arms. Our simulations show that such long filaments are even more prevalent and extended in the CO-dark molecular gas than in the visible molecular material. Therefore observations of CO-emitting filaments may probe just small parts of larger diffuse filaments many hundreds of parsecs in length.

\section{Simulation Comparison} \label{comparison}

\begin{figure*}
\begin{center}
\begin{tabular}{c c}
\includegraphics[width=3.5in]{./Fig_final/F2_ColDensity267_L035.eps}
\includegraphics[width=3.5in]{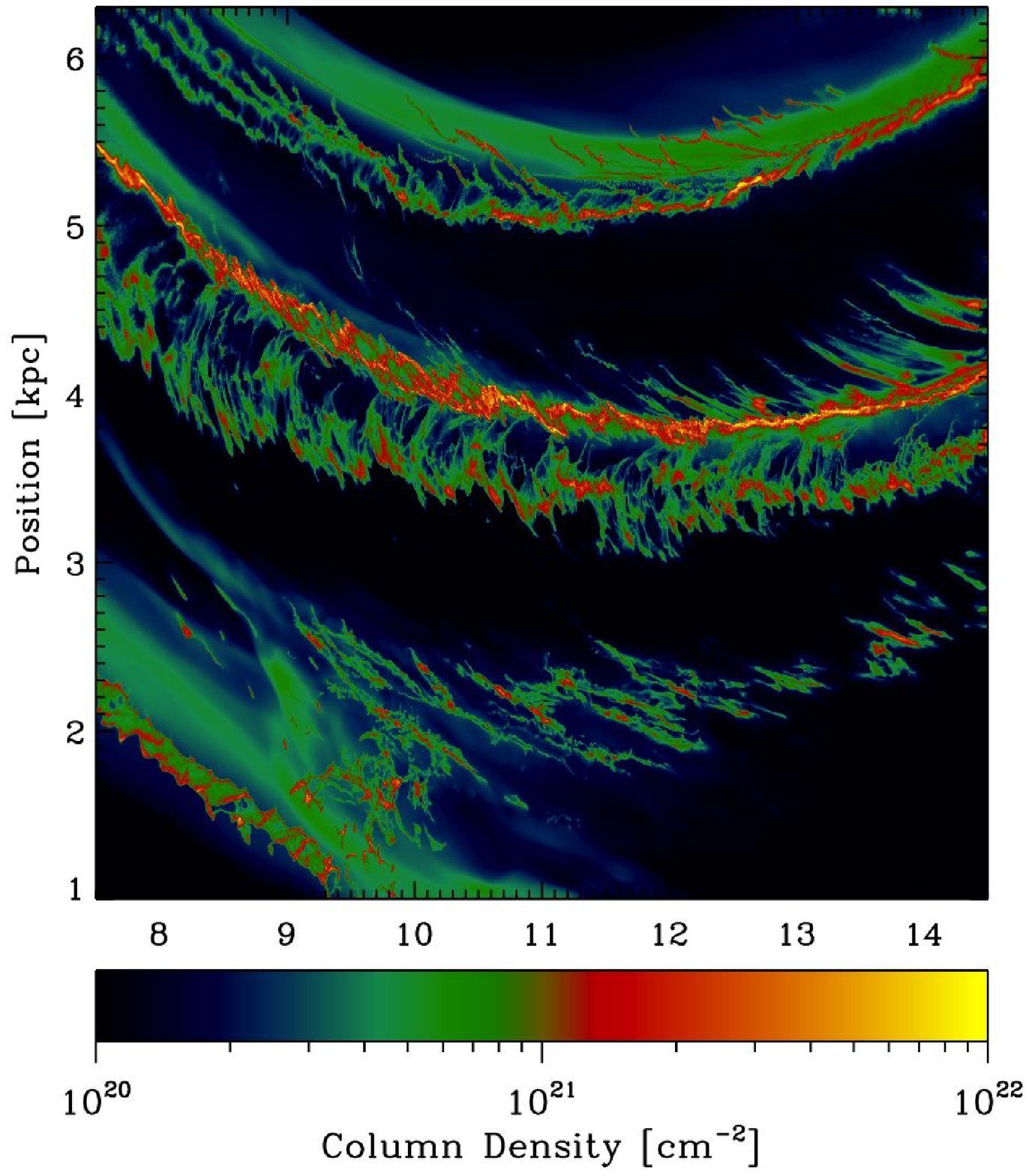}\\
\includegraphics[width=3.5in]{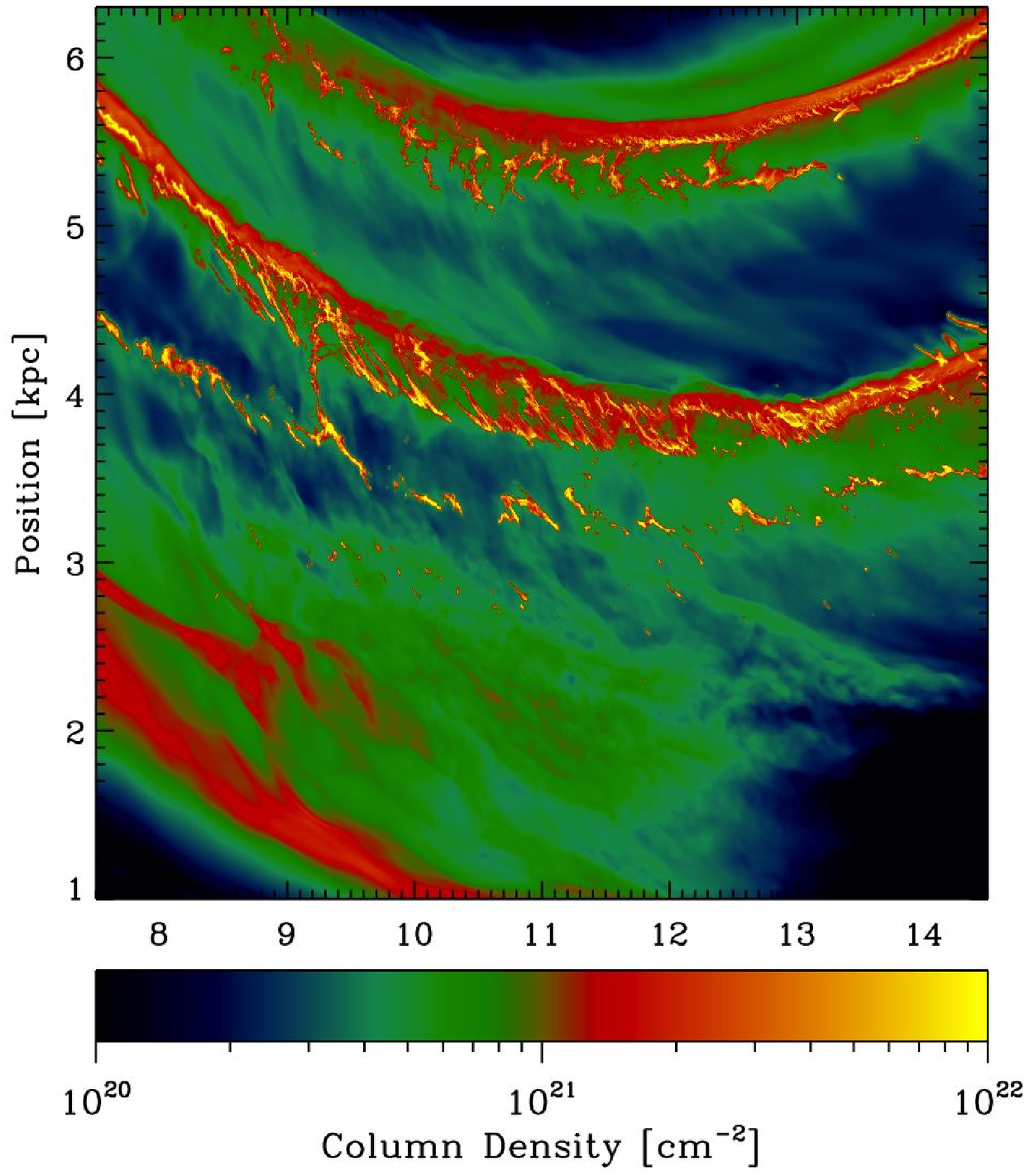}
\includegraphics[width=3.5in]{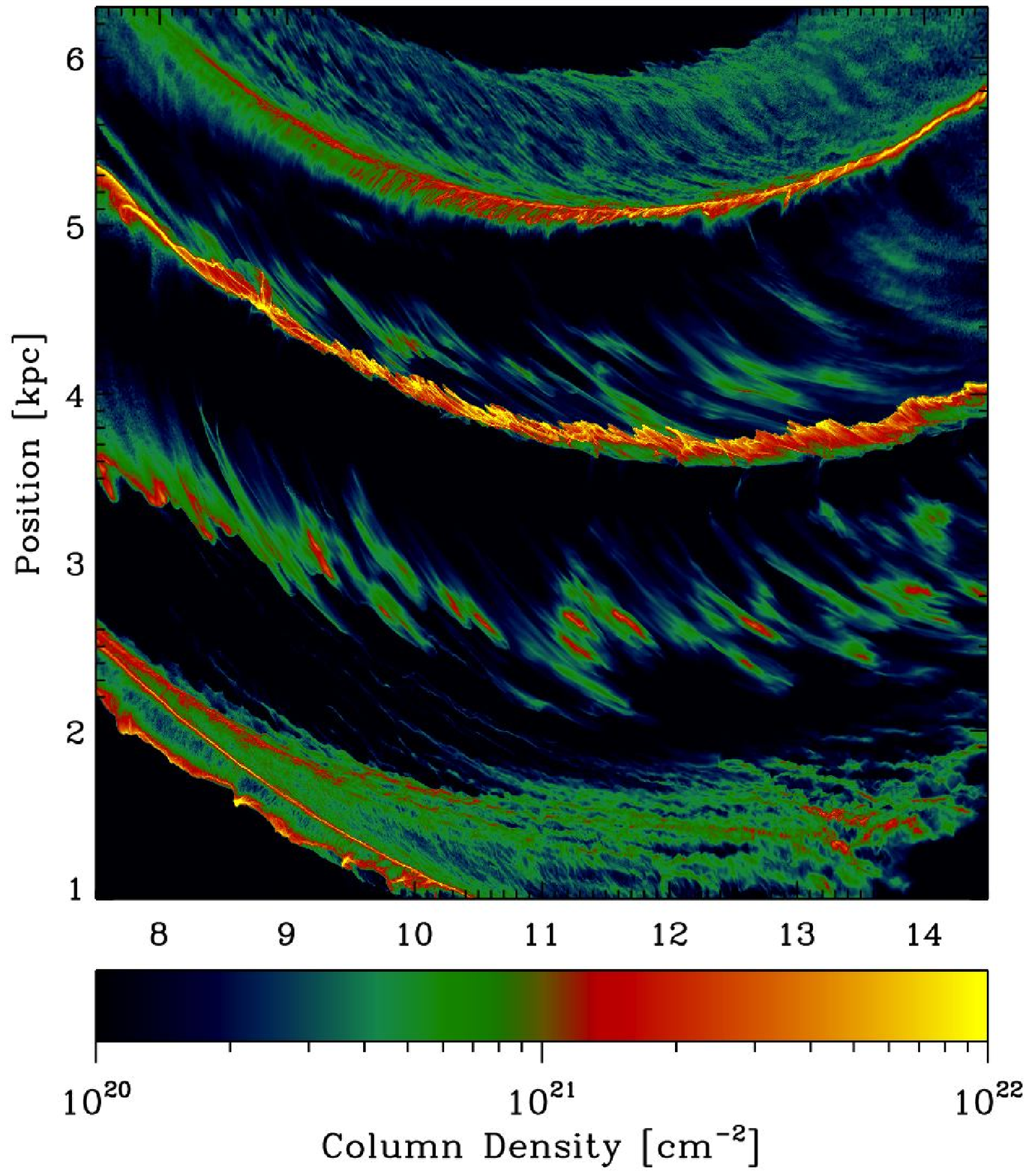}\\
\end{tabular}
\caption{Column density maps of the four simulations. \textit{Top left}: the fiducial Milky Way run. \textit{Top right}: the Low Density run. \textit{Bottom left}: the Strong Field run. \textit{Bottom right}: the Low \& Weak run. Both the average surface density of the disc and the strength of the ambient ISRF affect the morphology of the clouds in the disc.}
\label{comp_col}
\end{center}
\end{figure*}

\begin{figure*}
\begin{center}
\begin{tabular}{c c}
\includegraphics[width=3.5in]{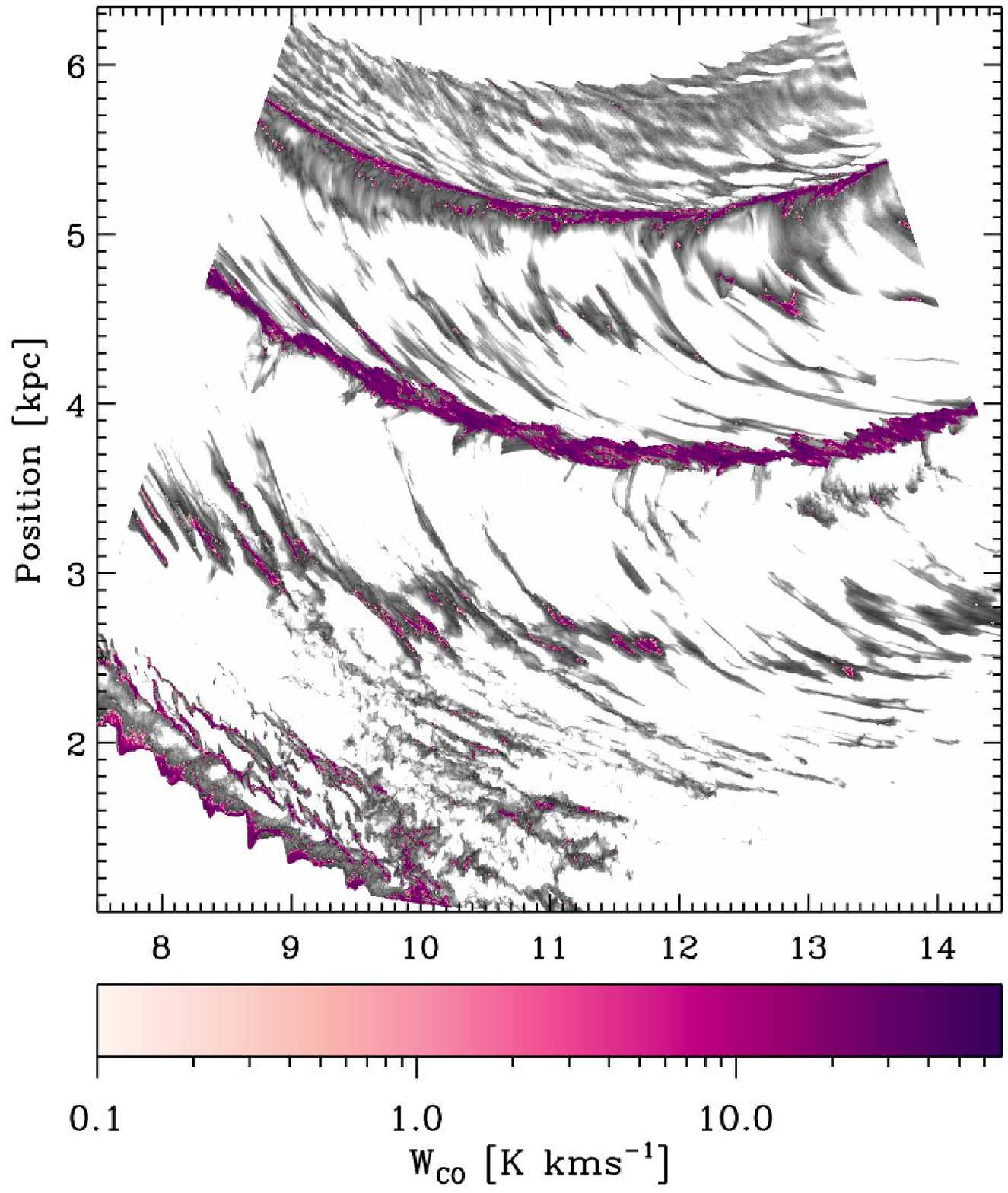}
\includegraphics[width=3.5in]{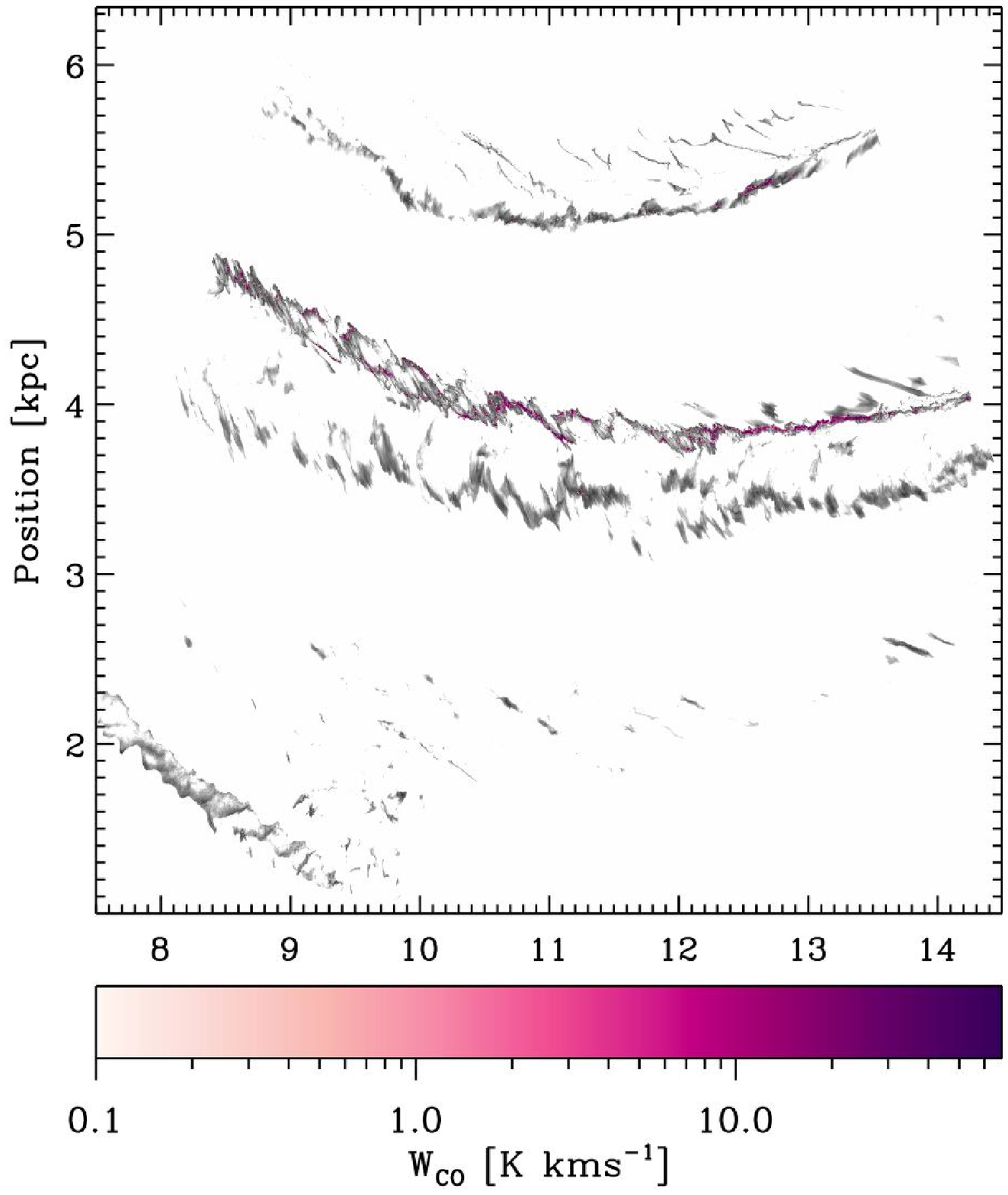}\\
\includegraphics[width=3.5in]{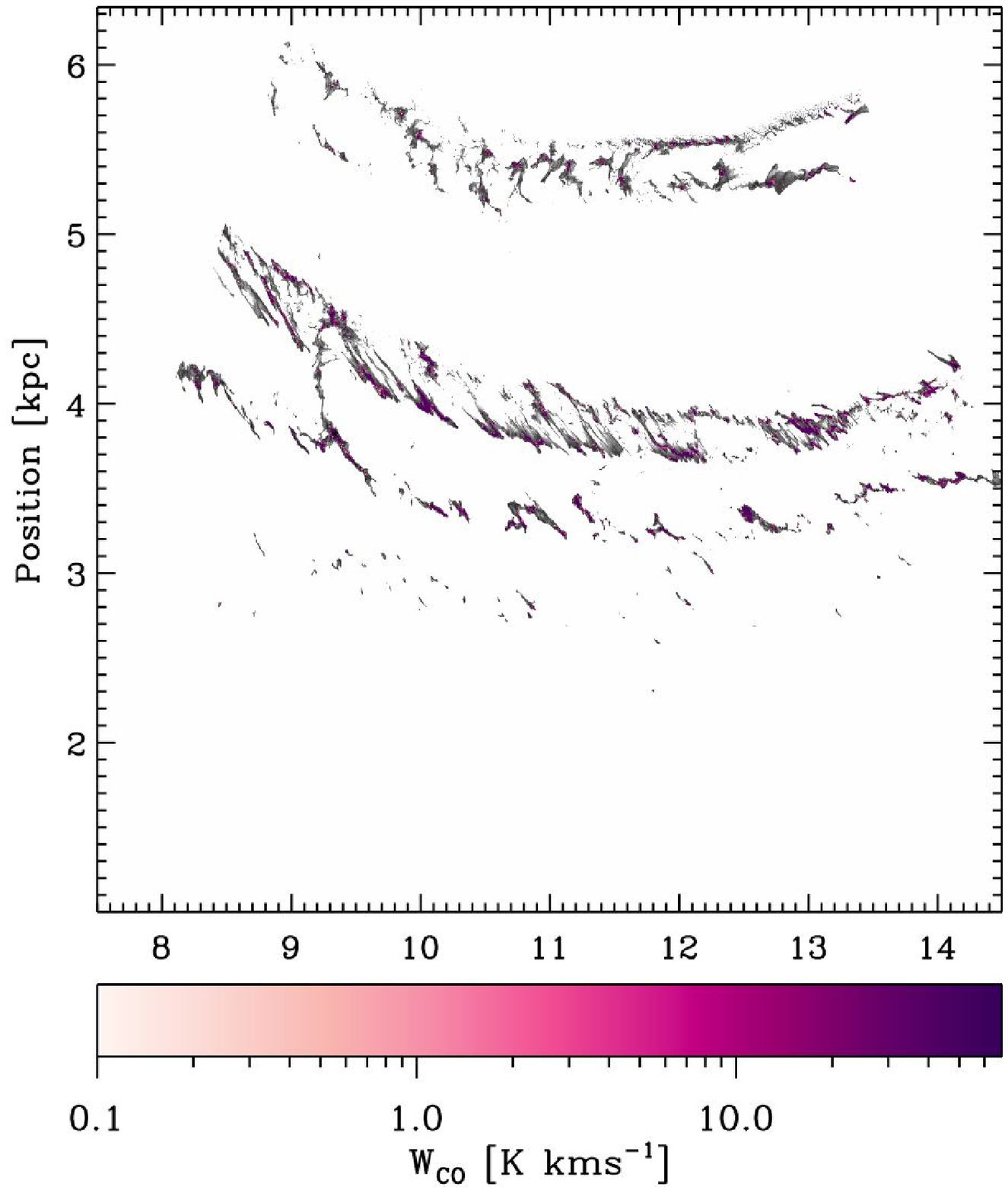}
\includegraphics[width=3.5in]{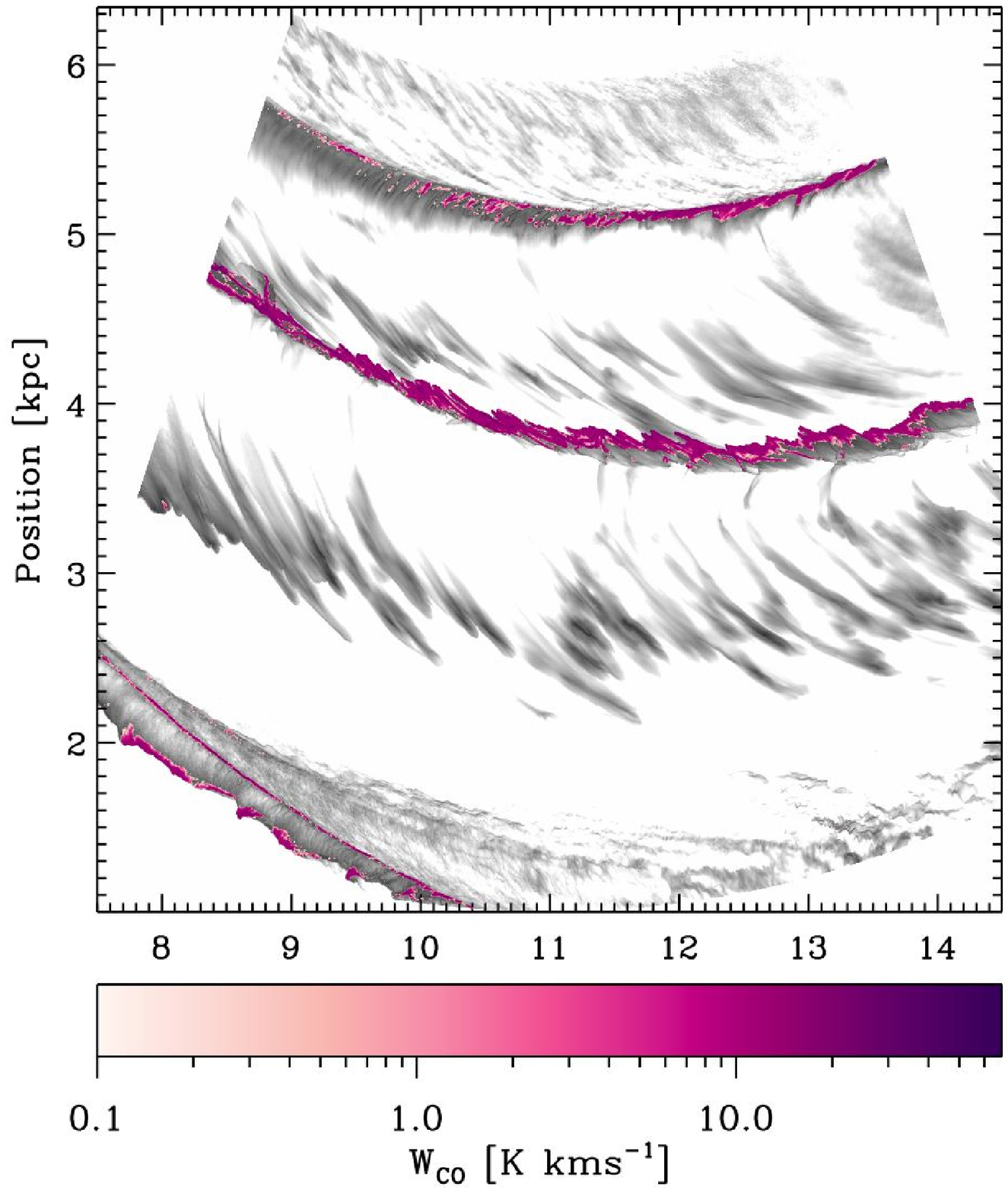}\\
\end{tabular}
\caption{Maps of the CO integrated intensity for the same four simulations as in \fig \ref{comp_col}.  (\textit{Top left}: the Milky Way run. \textit{Top right}: the Low Density run. \textit{Bottom left}: the Strong Field run. \textit{Bottom right}: the Low \& Weak run). The grayscale background image shows the H$_{2}$ column density, $N_{\rm H_{2}}$, as in Figure~\ref{morphology}. In the low surface density cases, there is essentially no CO emission from the inter-arm regions and very little H$_{2}$ is located there.}
\label{comp_wco}
\end{center}
\end{figure*}

Having examined in detail the Milky Way simulation we now turn our attention to the remaining simulations to see how they differ. In Figures \ref{comp_col} and \ref{comp_wco} we show maps of the total column density, the H$_{2}$ column density, and the CO integrated intensity for each of our four simulations. In the Low Density run, which has the same ISRF as our fiducial Milky Way run, but a mean gas surface density of only $4 \: {\rm M_{\odot}} \: {\rm pc^{-2}}$ (compared to $10 \: {\rm M_{\odot}} \: {\rm pc^{-2}}$ in the Milky Way run), significantly less H$_{2}$ is formed. The CO emission in this simulation is located almost entirely within the spiral arms, with little emission coming from the inter-arm regions. In the Low \& Weak simulation, in which the surface density is the same as in the Low Density run, but the strength of the ISRF is reduced by a factor of ten, we find a lot more diffuse H$_{2}$ and more bright CO emission from the spiral arms, but we still do not see significant emission from the inter-arm regions. Finally, in the Strong Field simulation, which has the same surface density as our Milky Way simulation, but an ISRF that is a factor of ten stronger, we find little H$_{2}$ or CO outside of the very densest gas. In this simulation, the transition between dense shielded material and the outside is very sharp and the diffuse structure of the disc has a smoother density distribution due to its generally higher temperature.

\begin{figure*}
\begin{center}
\includegraphics[width=3in]{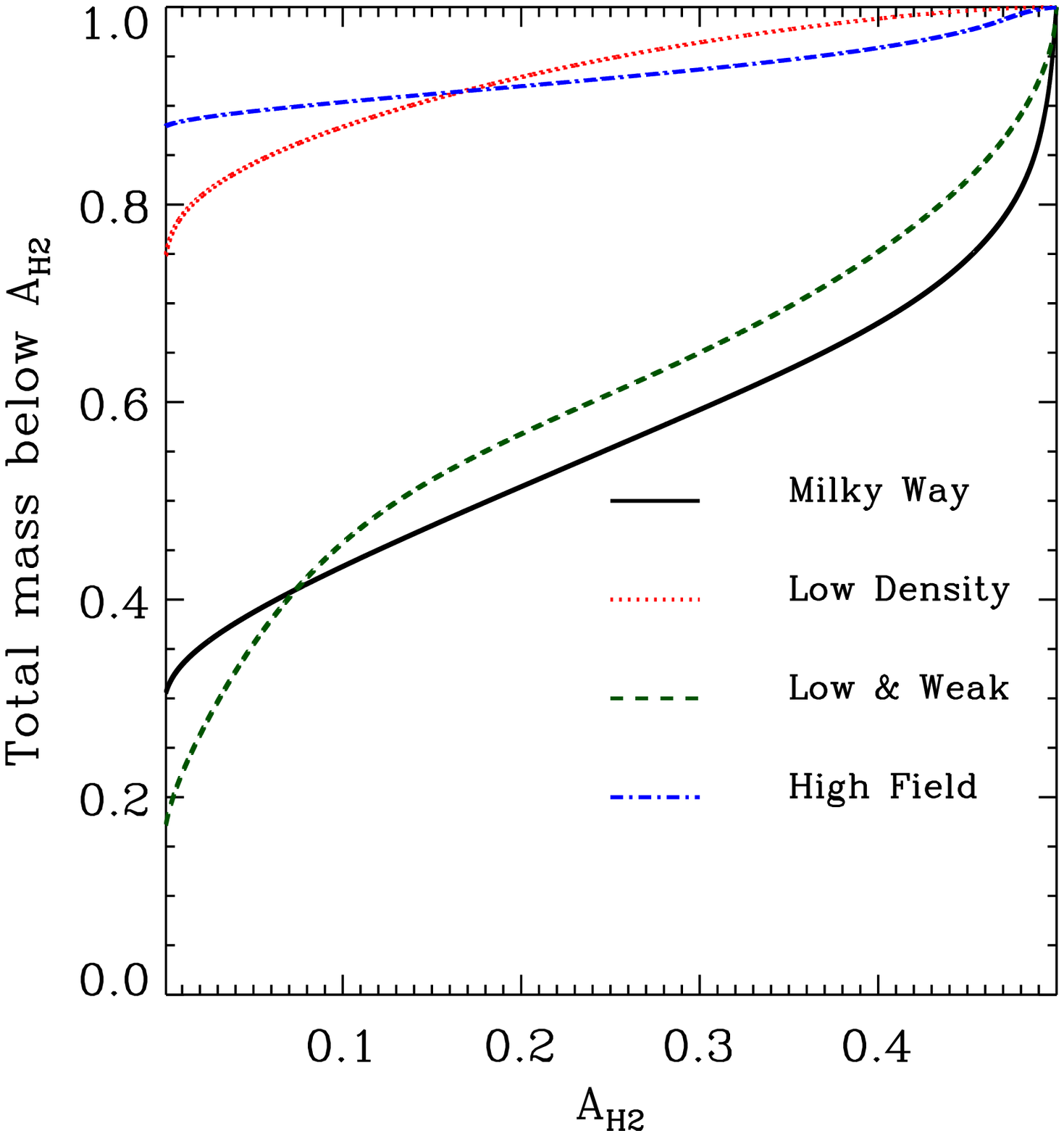}
\includegraphics[width=3in]{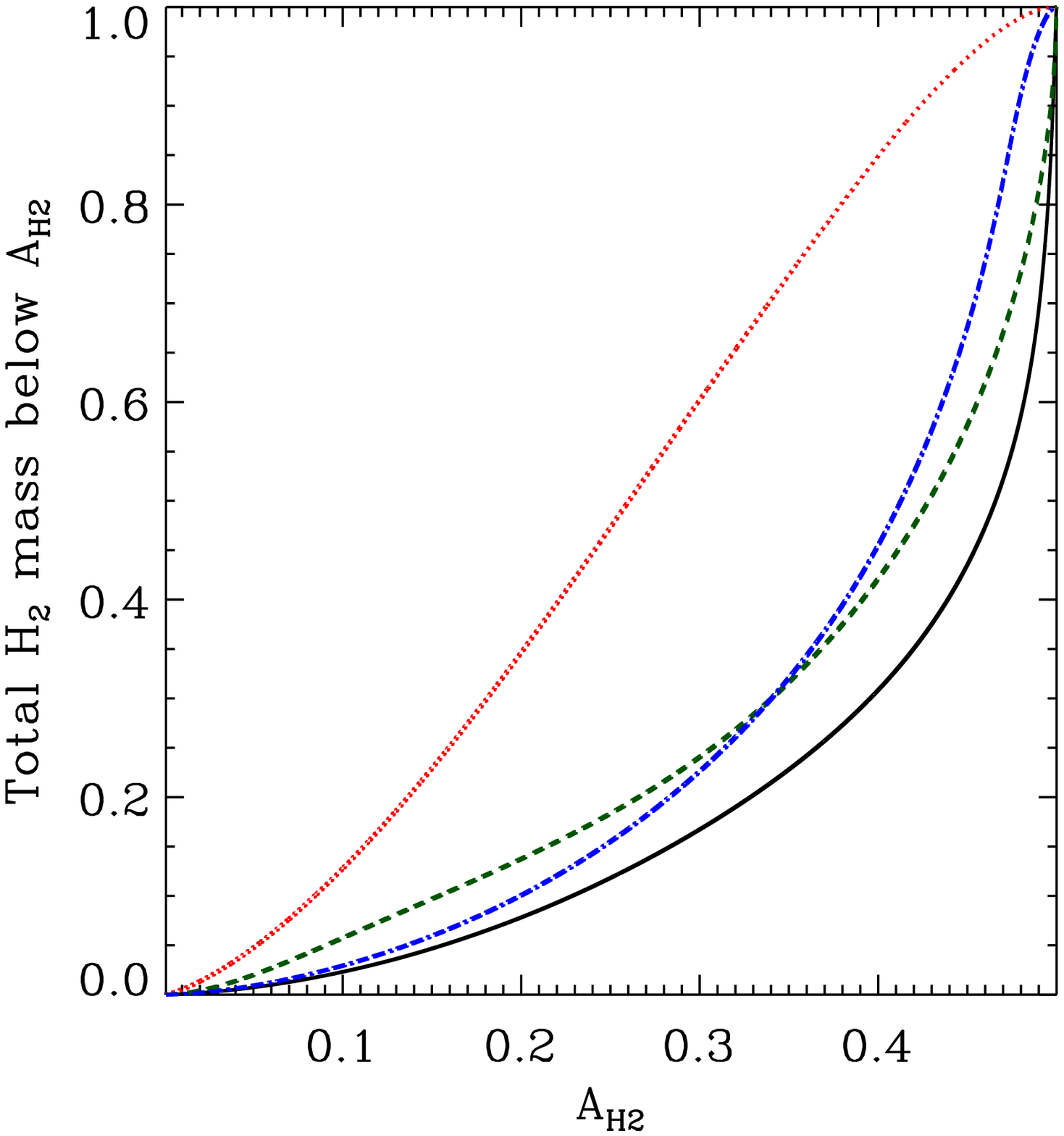}
\caption{The fraction of the total gas mass, and the total H$_2$ mass, in each simulation below a given H$_2$ abundance $A_{\rm H_{2}}$. In all the simulations a substantial fraction of the gas is partially molecular.}
\label{fh2_frac}
\end{center}
\end{figure*}

In Figure~\ref{fh2_frac}, we show how the total gas mass and the mass of H$_{2}$ are distributed as a function of the fractional H$_{2}$ abundance, $A_{\rm H_{2}}$. In our fiducial Milky Way simulation, we see that around half of the total mass is in molecular-dominated regions with $A_{\rm H_{2}} > 0.25$ (i.e.\ regions in which there is more H$_{2}$ than H{\sc i}). If we focus instead on the proportion of the total disc mass in a single phase we find that 43\% of the gas is predominantly atomic hydrogen ($A_{\rm H_{2}} < 0.1$), 32\% of the gas is predominantly molecular hydrogen ($A_{\rm H_{2}} > 0.1$), and 25\% has a mixture of atomic and molecular hydrogen. 
It is perhaps interesting to note that there is growing observational evidence for the presence of this mixed component in the real ISM, coming from {\em Herschel} observations of molecular ions such as OH$^{+}$ and H$_{2}$O$^{+}$ \citep[see e.g.][]{Neufeld10}. However, if we look at the distribution of the H$_{2}$ \textit{mass}, then we find, not unexpectedly, that a large fraction is located in regions with high $A_{\rm H_{2}}$. Around 50\% of the total molecular mass is located in regions with $A_{\rm H_{2}} > 0.45$ (corresponding to gas that is more than 90\% molecular), while only a little over 10\% is located in regions where the majority of the gas is atomic. 

Figure \ref{fh2_frac} also shows the gas mass distribution for our three comparison simulations. The High Field and Low Density simulations both have extremely high atomic gas fractions of $\sim$ 88\% and $\sim$ 75\% respectively due to the difficulty of forming H$_2$ in diffuse or highly irradiated gas. The Low \& Weak simulation has a similar behaviour for the total gas mass abundance as the Milky Way simulation, but has a slightly lower atomic gas fraction $\sim$ 20\%. In a similar manner to the Milky Way simulation most of the mass in molecular hydrogen is at high abundances in the High Field and Low \& Weak simulations. The exception to this behaviour is the Low Density simulation in which there is an almost linear increase in the total H$_2$ mass below a given $A_{\rm H_{2}}$. In all cases there is a non-negligible fraction of the gas which is only partially molecular. This partially molecular phase is typically at lower column densities than the fully molecular gas and therefore more likely to be CO-dark.

\begin{figure*}
\begin{center}
\includegraphics[width=3in]{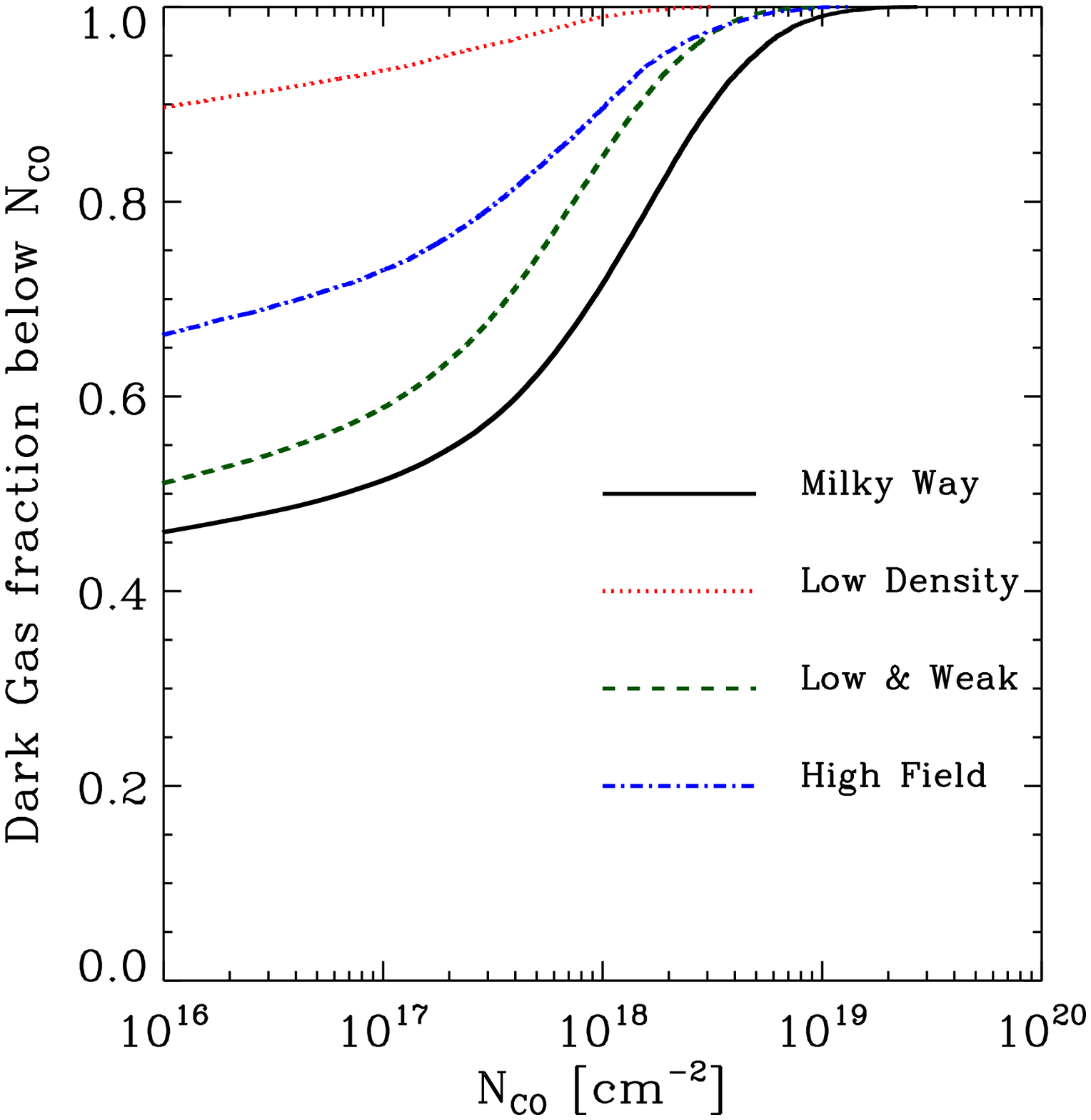}
\includegraphics[width=3in]{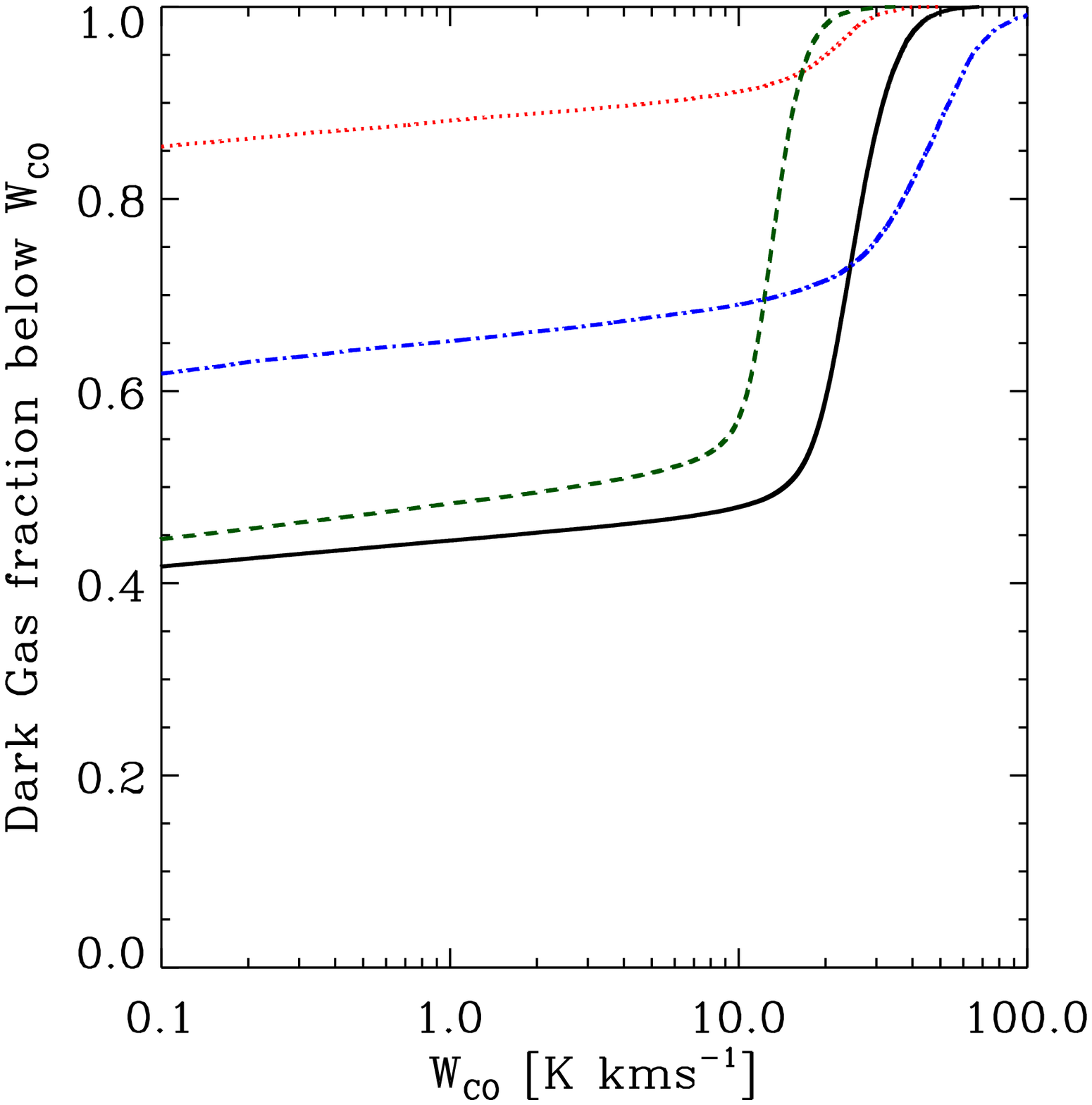}
\caption{Cumulative plots of H$_{2}$ mass as a function of $N_{\rm CO}$ ({\em left}) or $W_{\rm CO}$ ({\em right}) for our four simulations. The Milky Way simulation has the lowest dark gas fraction, which suggests that the dark gas fraction might be higher in external galaxies with a lower surface density or higher ambient radiation field.}
\label{comp_darkgas}
\end{center}
\end{figure*}

In \fig \ref{comp_darkgas}, we examine how the dark gas fraction varies as a function of $N_{\rm CO}$ and $W_{\rm CO}$ in our four simulations. We see that the Milky Way  simulation has the lowest dark gas fraction of the four simulations, regardless of whether we use a column density threshold or integrated intensity threshold to define the dark gas. In all four of the simulations, we see similar behaviour for $f_{\rm DG}$: it initially rises slowly with increasing $N_{\rm CO}$ or $W_{\rm CO}$, before suddenly rising more abruptly once we reach the conditions corresponding to CO-bright GMCs. Increasing the radiation field strength or decreasing the mean surface density of the disc makes it easier to photodissociate the CO and hence leads to a higher dark gas fraction. On the other hand, if we decrease the mean surface density but at the same time lower the radiation field strength, as in our Low \& Weak simulation, we see that the two effects largely cancel and that the dark gas fraction is not very different from the value in our fiducial Milky Way simulation. In  \tab \ref{Xco_tab}, we give the values of $f_{\rm DG}$ that we derive if we take a CO column density threshold $N_{\rm CO} = 10^{16} \: {\rm cm^{-2}}$ or an integrated intensity threshold $W_{\rm CO} = 0.1 \: {\rm K \, km \, s^{-1}}$. In the Milky Way, adopting this threshold means that we miss $\sim$45\% of the H$_{2}$, but we see that in other galaxies with lower surface densities or higher radiation fields, we can easily miss much more of the H$_{2}$.

Comparison of the results of the four runs also shows us that the dark gas fraction is far more sensitive to the mean surface density of the disc, and hence the characteristic surface density of the clouds that form in the disc, than it is to the strength of the interstellar radiation field. Changing the surface density by a factor of 2.5 has a greater effect on $f_{\rm DG}$ than changing the strength of the ISRF by a factor of ten. This behaviour has a simple explanation in terms of the microphysics of CO photodissociation. We can write the CO photodissociation rate as a function of the visual extinction of the gas as
\begin{equation}
R_{\rm pd} = R_{\rm pd, thin} \exp(-\gamma A_{\rm V}),
\end{equation}
where $R_{\rm pd, thin}$ is the CO photodissociation rate in optically thin gas, $A_{\rm V}$ is the visual extinction, $\gamma = 2.5$ for the treatment of CO photodissociation used in our simulations, and where, for simplicity, we have ignored the effects of CO self-shielding. Changing the strength of the ISRF changes $R_{\rm pd, thin}$ by the same amount and hence $R_{\rm pd}$ varies linearly with the radiation field strength. On the other hand, altering the mean surface density of the gas leads to changes in the mean visual extinction of the clouds that form, and hence can lead to exponential changes in $R_{\rm pd}$.

Finally, it is interesting to look at how the globally-averaged CO-to-H$_{2}$ conversion factor varies as we alter the simulation parameters. We list in \tab \ref{Xco_tab} the values of \XCO that we have calculated for all four of our simulations. As before, we calculate two values of \XCO, one for the gas where CO is visible (\WCO $> 0.1$ K km s$^{-1}$), and another for all of the gas (including the CO-dark molecular and atomic gas). In all cases the value of \XCO calculated using the entirety of the gas is higher than the value of \XCO using only the regions where there is visible CO emission. The discrepancy between the two values increases in the simulations with a greater dark gas fraction. When considering the value of \XCO in the CO-bright gas, \XCO is marginally lower than the Milky Way simulation in the Low Density simulation due to the lower column density of H$_2$, and \XCO is lower in the High Field simulation due to the brighter emission from hotter CO. In the Low \& Weak simulation the competing effects of a lower H$_2$ column density and lower CO gas temperature cancel each other out to give a similar value of \XCO to the Milky Way simulation.

\begin{table*}
	\centering
	\caption{Dark gas fractions and mean CO-to-H$_{2}$ conversion factors for our four simulations.}
		\begin{tabular}{l c c c c}
		\hline
	         \hline
	          Simulation & Dark Gas  & Dark Gas  & X$_{\rm CO}$ (CO bright) & X$_{\rm CO}$ (All gas) \\
	           & ($N_{\rm CO} \leq 10^{16} \, {\rm cm^{-2}}$) & ($W_{\rm CO} \leq 0.1\, {\rm K \, km \, s^{-1}}$) & [${\rm cm^{-2}\, K^{-1}\, km^{-1}\, s}$] &[${\rm cm^{-2}\, K^{-1}\, km^{-1}\, s}$] \\
	   	 \hline
		 Milky Way  & 46\%& 42\% &$2.28 \E^{20}$ & $3.91 \E^{20}$ \\
		 Low Density  & 90\% & 85\% &$1.92 \E^{20}$ & $1.31 \E^{21}$ \\
		 Strong Field & 66\% & 62\% & $1.48 \E^{20}$ & $3.89 \E^{20}$ \\
		 Low \& Weak & 51\% & 45\%& $2.20 \E^{20}$ & $3.96 \E^{20}$ \\
	         \hline	 
	         \hline        
		\end{tabular}
	\label{Xco_tab}
\end{table*}

\section{Discussion}\label{discussion}
\subsection{Comparison with previous work}

Previous theoretical work on quantifying the CO-dark molecular gas fraction has been limited to studies of individual clouds. \citet{Wolfire10} used PDR models of spherical clouds to determine how much H$_{2}$ could be located in the CO-dark cloud envelope. Using the same definition of the dark gas fraction as in our present study, they derived a value $f_{\rm DG} \simeq 0.3$, with little dependence on cloud mass or radiation field strength, for clouds with mean extinctions characteristic of observed GMCs. They also showed that $f_{\rm DG}$ for a given cloud is a strong function of the mean extinction of the cloud, rising dramatically as this decreases. This fact likely explains the difference between the dark gas fraction we measure in our fiducial simulation, $f_{\rm DG}=0.42$, and the \citet{Wolfire10} value of 0.3 for GMCs. In our Milky Way simulation, a significant amount of the total H$_{2}$ content of the gas is located in long filamentary clouds in the interarm regions that because of their geometry have low effective visual extinctions, and hence high dark gas fractions. When we globally average $f_{\rm DG}$, we include these clouds as well as the CO-bright GMCs, and hence naturally recover a higher value for $f_{\rm DG}$.
 
Another interesting difference between our results and the \citet{Wolfire10} results is that we find the globally-averaged $f_{\rm DG}$ to be sensitive to the radiation field strength, rising to $f_{\rm DG}=0.62$ in our Strong Field run, whereas \citet{Wolfire10} find little or no dependence of $f_{\rm DG}$ on the radiation field strength. We can understand this deviation if we compare the column density distributions in our Milky Way and Strong Field simulations (Figure~\ref{comp_col}). Increasing the radiation field strength leads to an increased photoelectric heating rate in diffuse gas, and hence changes the relative abundance of gas in the cold neutral medium and warm neutral medium phases. Additionally, we also include increased cosmic ray ionisation in our Strong Field case. Although direct heating of the diffuse gas by the cosmic rays is unimportant in comparison to photoelectric heating, the increased ionisation that they provide increases the photoelectric heating efficiency, further boosting the amount of warm gas. As a result, fewer massive dense clouds form in our Strong Field simulation. Therefore, although the dark gas fraction of these dense clouds probably does not change significantly with the increase in radiation field strength (see also Clark \& Glover, in prep.), the global average does change, as more of the H$_{2}$ is now located in lower density, low $A_{\rm V}$ clouds with high dark gas fractions. 

Observationally, several attempts have been made to constrain $f_{\rm DG}$. For example, \citet{Grenier05} used large-scale maps of H{\sc i} and CO emission, dust emission and extinction, and $\gamma$-ray emission (which traces the total gas column density) to show that there was a significant fraction of gas that was traced in dust emission and absorption and in $\gamma$-ray emission, but that was not seen in either H{\sc i} or CO emission. They estimated the total mass in this dark fraction to be 50--100\% of the mass in CO-bright clouds. Adopting our definition of $f_{\rm DG}$, this corresponds to values in the range $f_{\rm DG}=0.33$ to $f_{\rm DG}=0.5$, in good agreement with the value of $f_{\rm DG}=0.42$ that we find in our Milky Way simulation. 

The Planck collaboration also investigated the dark gas fraction by constructing an all-sky map of dust temperature and optical depth and correlating this with H{\sc i} and CO emission. An excess in dust emission compared to these tracers was interpreted as belonging to CO-dark gas, with a mass at high Galactic latitudes that was 118\% times that of the CO-bright gas \citep{Planck11}. Using our definition of $f_{\rm DG}$ (which differs from that used in their paper), this corresponds to  $f_{\rm DG} = 0.54$, somewhat higher than the value we derive from the Milky Way system. However, a direct comparison with our results is hampered by the fact that they do not consider clouds close to the Galactic mid-plane, which we might reasonably expect to have higher surface densities than high-latitude clouds. Another dust-based study was carried out by \citet{Paradis12}, but using dust extinction maps instead of the Planck far-infrared maps. They found an even higher dark gas fraction of $f_{\rm DG}=0.62$, but again considered only high Galactic latitudes ($|b| > 10^{\circ}$). They also showed that the dark gas fraction appeared to be significantly larger in the inner Galaxy ($f_{\rm DG}=0.71$) than in the outer Galaxy ($f_{\rm DG} = 0.43$).

\begin{figure}
\begin{center}
\includegraphics[width=3.2in]{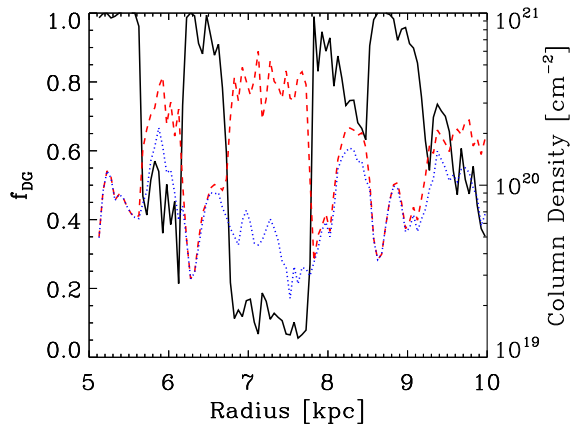}
\caption{Dark gas fraction, $f_{\rm DG}$, as a function of galactocentric radius in the Milky Way simulation (solid line). The red dashed line shows the mean molecular hydrogen column density \NHm at the corresponding radius and the blue dotted line shows the contribution to \NHm from gas that is CO-dark. The dark gas fraction is anti-correlated with \NHm.}
\label{radial}
\end{center}
\end{figure}

One drawback of these observations is that they provide information only on the projected (i.e.\ line-of-sight averaged) value of $f_{\rm DG}$. However, as Figure~\ref{morphology} makes plain, we expect there to be considerable spatial structure present in the dark gas distribution: the value will be much higher in inter-arm regions and consequently will be smaller than the globally-averaged value inside the spiral arms. We can see this more clearly if we plot $f_{\rm DG}$ as a function of radius for our Milky Way simulation (\fig \ref{radial}). In the low surface density regimes, \NHm and the dark gas surface density are almost identical and the dark gas fraction can be of order unity. At higher column densities, the dark gas fraction drops:  for example, the spiral arm situated at a radius of around 7~kpc only has a dark gas fraction of between 10 and 20 percent. In general the dark gas fraction is anti-correlated with \NHm. A similar finding that the fraction of dark gas is higher in diffuse molecular clouds has also been found observationally by \citet{Langer14}.

An observational approach that allows one to look for structures in the distribution of $f_{\rm DG}$ involves using [C{\sc ii}] emission as a tracer of CO-dark gas. \citet{PinedaJ13} use the GOT~C+ (Galactic Observations of Terahertz C$^+$) survey \citep{Langer10} to study the distribution of [C{\sc ii}] emission in the Milky Way. This survey consists of 500 evenly distributed lines-of-sight in the Galactic plane that were observed at 158$\, \mu$m using the HIFI instrument \citep{Hifi2010} on board the {\em Herschel} space telescope \citep{Pilbratt10}. By correlating the detected [C{\sc ii}] emission with complementary maps of H{\sc i} and CO emission, \citet{PinedaJ13} were able to identify regions with excess [C{\sc ii}] emission that could not be explained by emission from atomic clouds or CO-bright molecular clouds. They interpreted this excess emission as coming from clouds of CO-dark molecular gas.

Averaged over their whole survey, \citet{PinedaJ13} find 
a dark gas fraction $f_{\rm DG} \simeq 0.30$. However, they also show that $f_{\rm DG}$ varies substantially with Galactocentric distance: within $R_{\rm G} = 3 \: {\rm kpc}$ of the Galactic centre it is essentially zero, whereas by the time we reach $R_{\rm G} = 10 \: {\rm kpc}$, $f_{\rm DG} \simeq 0.6$--0.8. They hypothesise that this trend is due to the influence of the Galactic metallicity gradient -- clouds with similar column densities in the inner and outer Galaxy will have different mean extinctions and hence different dark gas fractions. Nevertheless, it should be noted that their finding of a low $f_{\rm DG}$ in the inner Galaxy and a high $f_{\rm DG}$ in the outer Galaxy seems to be in contradiction with the results of \citet{Paradis12} discussed above, who find the opposite. \citet{PinedaJ13} also present some evidence for smaller-scale variations in $f_{\rm DG}$, although these variations do not appear to be nearly as large as those we see in Figure~\ref{radial}. One possible explanation for why we see much larger variations in $f_{\rm DG}$ than are seen in the real data is that we assume that the strength of the interstellar radiation field is constant throughout our simulated disc. In reality, one would expect it to be stronger in the spiral arms, close to the sites of massive star formation, and weaker in the inter-arm regions. That said, it is also possible that the \citet{PinedaJ13} measurements may systematically underestimate the dark gas fraction in inter-arm regions if a significant fraction of the dark gas is too cool to emit in [C{\sc ii}]. 

In summary, the results of our Milky Way simulation appear to agree reasonably well with local determinations of $f_{\rm DG}$ and with previous theoretical work. We do not recover the correlation between $f_{\rm DG}$ and Galactocentric radius reported by \citet{PinedaJ13}, but as our disc has a constant metallicity and constant mean surface density, unlike the real Milky Way, we would not expect to be able to reproduce this feature. 

\subsection{Sources of Uncertainty}
There are a number of potential uncertainties in our simulations. As we have already discussed, we assume a uniform interstellar radiation field, but this assumption is obviously false on small scales, particularly close to regions of massive star formation. On large scales, however, this should be a reasonable approximation provided that we consider a region within a few kpc of the Solar Galactocentric radius \citep{Wolfire03}. Another simplification is that we have assumed a constant shielding length of 30 pc between the gas and the radiation source in all the simulations. If the star formation density were constant, one could suppose that this distance should be smaller for higher field strengths, and larger for lower field strengths. Alternatively, if the star formation is primarily clustered, and the cluster mass increases with star formation rate, a constant shielding length would remain a reasonable approximation. Even in the Milky Way there may be some cause to doubt this assumption, as the shielding length might be shorter in the spiral arms. Since these simulations are too computationally expensive to do a full parameter study, our constant shielding length represents a reasonable first estimate of what the local shielding may be.

We also assume a constant metallicity and dust-to-gas ratio, whereas the Milky Way galaxy has a slight metallicity gradient with Galactocentric radius. \citet{Luck11} find that the iron distribution can be fit by the relationship $[{\rm Fe/H}] = -0.055 R_{\rm G} + 0.475$ in dex, where $R_{\rm G}$ is the Galactocentric radius in kpc. For the radii included in our model, this equates to a range of 0.2 to -0.075 dex which does not substantially fall below the assumed solar metallicity. In future works we plan to address how metallicity affects the dark gas fraction in more detail.

Further uncertainties stem from the fact that our model is an idealised treatment of the galaxy which uses an external potential to produce tightly wound spiral arms and does not include the effects of gas self-gravity or stellar feedback. We expect the effects of these additional processes to cancel out to some extent, as a plausible explanation for why the star formation efficiency of the Galaxy is so low is that the energy input from stellar feedback maintains molecular clouds in a gravitationally unbound state \citep{Dobbs11}.

Finally, some additional uncertainty is introduced by the fact that our treatment of the CO chemistry is approximate and tends to under-predict the actual CO abundance along very low column density sight-lines, and by our equally approximate method for determining \WCO. However, as we have already discussed in some detail in Section \ref{xfactor}, neither of these uncertainties are likely to strongly affect the results that we derive for $f_{\rm DG}$, because the sharp increase in the CO emission from CO-bright compared to CO-dark gas is far larger than our uncertainties (see Figure~\ref{darkWCO}). 

\section{Conclusions} \label{conclusions}
We have performed high-resolution simulations of a significant portion of a galactic disc with a version of the \arepo moving mesh code that includes a self-consistent treatment of the chemistry of the ISM in order to investigate the fractional amount and morphology of CO-dark molecular gas. In the most highly refined portion of our simulation we reach a mass resolution of only four solar masses. To the best of our knowledge, this is the highest resolution treatment to date of the ISM on galactic scales that has followed the chemical evolution of the gas. This approach enables us to resolve substructure within the disc sufficiently accurately to match the observed H{\sc i}--H$_2$ transition without having to resort to an ad-hoc ``clumping factor'' to represent unresolved density fluctuations.

We considered four different models: a fiducial `Milky Way' model designed to be a reasonable match to the properties of our own Galaxy in the solar neighbourhood, plus three comparison simulations in which we explored the effects of varying the mean surface density of the gas and the strength of the interstellar radiation field. We evolved the simulations for 261.1 Myr (corresponding to six spiral arm passages) and then calculated the amount and morphology of CO-bright and CO-dark molecular gas. Our major findings are outlined below:

\begin{enumerate}

\item In conditions typical of the Milky Way disc, 46\% of the molecular gas resides below CO column densities of $10^{16}$ \cms. In terms of emission, 42\% of the molecular gas  is found in regions with integrated intensities in the $J = 1 \rightarrow 0$ line of $^{12}$CO that are less than \WCO = 0.1~K~kms$^{-1}$. If we take this integrated intensity to mark the boundary between CO-bright and CO-dark gas, then we derive a dark gas fraction $f_{\rm DG} = 0.42$. Changing the value of our intensity threshold by factors of a few changes this number by only a few percent.

\item Our predicted dark gas fraction of $f_{\rm DG} = 0.42$ is in excellent agreement with estimates derived from gamma rays, slightly lower than estimates from dust emission and absorption, and is slightly higher than estimates derived from [C{\sc ii}] emission. Overall our estimate of the CO-dark gas fraction is consistent with the observational literature.

\item Although some CO-dark molecular gas is found in the vicinity of CO-bright clouds located in spiral arms, there is also a substantial amount in inter-arm regions in the form of extremely long filamentary clouds, with lengths of hundreds of parsecs. This filamentary geometry makes the CO more susceptible to dissociation due to the increased surface-to-volume ratio making it more difficult to shield. Observations have shown that filamentary molecular clouds are a feature of the Milky Way disc, but our simulations suggest that these might be just the observable parts of much larger structures.

\item A non-negligible fraction of the total disc mass exists in a mixed phase that is neither fully atomic nor fully molecular. In the Milky Way simulation we find that 25\% of the gas has an H$_2$ abundance relative to the total number density of hydrogen nuclei of between 0.1 and 0.4 (where a value of 0.5 corresponds to fully molecular gas). This partially molecular phase is typically at lower column densities than the fully molecular gas and therefore is more likely to be CO-dark. However the majority of the H$_2$ mass still resides in regions where the fractional abundance is greater than 0.45.

\item In simulations with higher radiation fields or lower mean surface densities, the dark gas fraction is much higher. However, the dark gas is still located primarily in the inter-arm regions of the disc. Higher radiation fields reduce the total amount of molecular gas and truncate the length of the filaments. Changes in the surface density of the disc have a much larger effect on $f_{\rm DG}$ than changes in the radiation field. These results suggest that external galaxies with lower surface densities or more intense radiation fields may have a greater dark gas fraction than the Milky Way. 

\item The globally-averaged CO-to-H$_2$ conversion factor that we measure in the CO visible gas in the Milky Way simulation, $X_{\rm CO} = 2.2\E^{20}$ cm$^{-2}$K$^{-1}$km$^{-1}$s, is in good agreement with the value inferred for the actual Milky Way. \XCO increases when the column density of the disc is reduced or the radiation field is increased, due to lower H$_2$ columns and brighter CO emission respectively. In all cases the value of \XCO rises if the diffuse gas is included in the estimate.

\end{enumerate}

\section*{Acknowledgements}
We acknowledge support from the following grants without which this work would not have been possible. R.J.S.\ and P.C.C.\ acknowledge support from the German Science Foundation (DFG) priority program 1573 {\em Physics of the Interstellar Medium} via projects SM 321/1-1 and CL 463/2-1, respectively. R.S.K.\ acknowledges support from the European Research Council under the European CommunityÕs Seventh Framework Programme (FP7/2007-2013) via the ERC Advanced Grant {\em STARLIGHT} (project number 339177). V.S.\ acknowledges support by the European Research Council under ERC-StG grant EXAGAL-308037. The research leading to these results have also been supported by the DFG via the Collaborative Research Centre SFB 881 {\em The Milky Way System} (sub-projects B1, B2, B5 and B8).

\bibliography{Bibliography}

\begin{thebibliography}{}

\bibitem[\protect\citeauthoryear{Abdo et~al.,}{Abdo  et~al.}{2010}]{Abdo10}
Abdo A.~A.,  et~al., 2010, \apj, 710, 133

\bibitem[\protect\citeauthoryear{{Adams}}{{Adams}}{1979}]{Adams79}
{Adams} D.~N.,  1979, {The Hitchhiker's Guide to the Galaxy}.
Pan Books

\bibitem[\protect\citeauthoryear{Ade et~al.,}{Ade  et~al.}{2011}]{Ade11}
Ade P.~A.~R.,  et~al., 2011, \aap, 536, 19

\bibitem[\protect\citeauthoryear{{Allen}, {Ivette Rodr{\'{\i}}guez}, {Black} \&
  {Booth}}{{Allen} et~al.}{2012}]{Allen2012}
{Allen} R.~J.,  {Ivette Rodr{\'{\i}}guez} M.,  {Black} J.~H.,    {Booth} R.~S.,
   2012, \aj, 143, 97

\bibitem[\protect\citeauthoryear{{Ballesteros-Paredes}}{{Ballesteros-Paredes}}{2006}]{Ballesteros06}
{Ballesteros-Paredes} J.,  2006, \mnras, 372, 443

\bibitem[\protect\citeauthoryear{{Bauer} \& {Springel}}{{Bauer} \&
  {Springel}}{2012}]{Bauer12}
{Bauer} A.,  {Springel} V.,  2012, \mnras, 423, 2558

\bibitem[\protect\citeauthoryear{{Bergin} \& {Tafalla}}{{Bergin} \&
  {Tafalla}}{2007}]{Bergin07}
{Bergin} E.~A.,  {Tafalla} M.,  2007, \araa, 45, 339

\bibitem[\protect\citeauthoryear{{Binney} \& {Tremaine}}{{Binney} \&
  {Tremaine}}{1987}]{Binney87}
{Binney} J.,  {Tremaine} S.,  1987, {Galactic Dynamics}.
Princeton University Press, Princeton, NJ

\bibitem[\protect\citeauthoryear{{Bolatto}, {Wolfire} \& {Leroy}}{{Bolatto}
  et~al.}{2013}]{Bolatto13}
{Bolatto} A.~D.,  {Wolfire} M.,    {Leroy} A.~K.,  2013, \araa, 51, 207

\bibitem[\protect\citeauthoryear{{Bonnell}, {Dobbs} \& {Smith}}{{Bonnell}
  et~al.}{2013}]{Bonnell13}
{Bonnell} I.~A.,  {Dobbs} C.~L.,    {Smith} R.~J.,  2013, \mnras, 430, 1790

\bibitem[\protect\citeauthoryear{{Burgh}, {France} \& {McCandliss}}{{Burgh}
  et~al.}{2007}]{Burgh07}
{Burgh} E.~B.,  {France} K.,    {McCandliss} S.~R.,  2007, \apj, 658, 446

\bibitem[\protect\citeauthoryear{{Caldwell} \& {Ostriker}}{{Caldwell} \&
  {Ostriker}}{1981}]{Caldwell87}
{Caldwell} J.~A.~R.,  {Ostriker} J.~P.,  1981, \apj, 251, 61

\bibitem[\protect\citeauthoryear{{Christensen}, {Quinn}, {Governato}, {Stilp},
  {Shen} \& {Wadsley}}{{Christensen} et~al.}{2012}]{Christensen12}
{Christensen} C.,  {Quinn} T.,  {Governato} F.,  {Stilp} A.,  {Shen} S.,
  {Wadsley} J.,  2012, \mnras, 425, 3058

\bibitem[\protect\citeauthoryear{{Clark}, {Glover} \& {Klessen}}{{Clark}
  et~al.}{2012}]{Clark12b}
{Clark} P.~C.,  {Glover} S.~C.~O.,    {Klessen} R.~S.,  2012, \mnras, 420, 745

\bibitem[\protect\citeauthoryear{{Clark}, {Glover}, {Klessen} \&
  {Bonnell}}{{Clark} et~al.}{2012}]{Clark12}
{Clark} P.~C.,  {Glover} S.~C.~O.,  {Klessen} R.~S.,    {Bonnell} I.~A.,  2012,
  \mnras, 424, 2599

\bibitem[\protect\citeauthoryear{{Contreras}, {Rathborne} \&
  {Garay}}{{Contreras} et~al.}{2013}]{Contreras13}
{Contreras} Y.,  {Rathborne} J.,    {Garay} G.,  2013, \mnras, 433, 251

\bibitem[\protect\citeauthoryear{{Cox} \& {G{\'o}mez}}{{Cox} \&
  {G{\'o}mez}}{2002}]{Cox02}
{Cox} D.~P.,  {G{\'o}mez} G.~C.,  2002, \apjs, 142, 261

\bibitem[\protect\citeauthoryear{{de Graauw} et~al.,}{{de Graauw}
  et~al.}{2010}]{Hifi2010}
{de Graauw} T.,  et~al., 2010, \aap, 518, L6

\bibitem[\protect\citeauthoryear{Dobbs}{Dobbs}{2006}]{Dobbs06}
Dobbs C.~L.,  2006, Ph.D Thesis, pp 20--36

\bibitem[\protect\citeauthoryear{{Dobbs} \& {Bonnell}}{{Dobbs} \&
  {Bonnell}}{2006}]{Dobbs06a}
{Dobbs} C.~L.,  {Bonnell} I.~A.,  2006, \mnras, 367, 873

\bibitem[\protect\citeauthoryear{{Dobbs}, {Burkert} \& {Pringle}}{{Dobbs}
  et~al.}{2011}]{Dobbs11}
{Dobbs} C.~L.,  {Burkert} A.,    {Pringle} J.~E.,  2011, \mnras, 413, 2935

\bibitem[\protect\citeauthoryear{{Dobbs}, {Glover}, {Clark} \&
  {Klessen}}{{Dobbs} et~al.}{2008}]{Dobbs08a}
{Dobbs} C.~L.,  {Glover} S.~C.~O.,  {Clark} P.~C.,    {Klessen} R.~S.,  2008,
  \mnras, 389, 1097

\bibitem[\protect\citeauthoryear{{Draine}}{{Draine}}{1978}]{Draine78}
{Draine} B.~T.,  1978, \apjs, 36, 595

\bibitem[\protect\citeauthoryear{{Draine} \& {Bertoldi}}{{Draine} \&
  {Bertoldi}}{1996}]{Draine96}
{Draine} B.~T.,  {Bertoldi} F.,  1996, \apj, 468, 269

\bibitem[\protect\citeauthoryear{{Federman}, {Huntress} Jr. \&
  {Prasad}}{{Federman} et~al.}{1990}]{Federman90}
{Federman} S.~R.,  {Huntress} Jr. W.~T.,    {Prasad} S.~S.,  1990, \apj, 354,
  504

\bibitem[\protect\citeauthoryear{{Gillmon}, {Shull}, {Tumlinson} \&
  {Danforth}}{{Gillmon} et~al.}{2006}]{Gillmon06}
{Gillmon} K.,  {Shull} J.~M.,  {Tumlinson} J.,    {Danforth} C.,  2006, \apj,
  636, 891

\bibitem[\protect\citeauthoryear{{Glover} \& {Clark}}{{Glover} \&
  {Clark}}{2012a}]{Glover12a}
{Glover} S.~C.~O.,  {Clark} P.~C.,  2012a, \mnras, 421, 116

\bibitem[\protect\citeauthoryear{{Glover} \& {Clark}}{{Glover} \&
  {Clark}}{2012b}]{Glover12b}
{Glover} S.~C.~O.,  {Clark} P.~C.,  2012b, \mnras, 421, 9

\bibitem[\protect\citeauthoryear{{Glover}, {Federrath}, {Mac Low} \&
  {Klessen}}{{Glover} et~al.}{2010}]{Glover10}
{Glover} S.~C.~O.,  {Federrath} C.,  {Mac Low} M.,    {Klessen} R.~S.,  2010,
  \mnras, 404, 2

\bibitem[\protect\citeauthoryear{{Glover} \& {Mac Low}}{{Glover} \& {Mac
  Low}}{2007a}]{Glover07a}
{Glover} S.~C.~O.,  {Mac Low} M.,  2007a, \apjs, 169, 239

\bibitem[\protect\citeauthoryear{{Glover} \& {Mac Low}}{{Glover} \& {Mac
  Low}}{2007b}]{Glover07b}
{Glover} S.~C.~O.,  {Mac Low} M.-M.,  2007b, \apj, 659, 1317

\bibitem[\protect\citeauthoryear{{Glover} \& {Mac Low}}{{Glover} \& {Mac
  Low}}{2011}]{Glover11}
{Glover} S.~C.~O.,  {Mac Low} M.-M.,  2011, \mnras, 412, 337

\bibitem[\protect\citeauthoryear{{Gnedin}, {Tassis} \& {Kravtsov}}{{Gnedin}
  et~al.}{2009}]{Gnedin09}
{Gnedin} N.~Y.,  {Tassis} K.,    {Kravtsov} A.~V.,  2009, \apj, 697, 55

\bibitem[\protect\citeauthoryear{{G{\'o}rski}, {Hivon}, {Banday}, {Wandelt},
  {Hansen}, {Reinecke} \& {Bartelmann}}{{G{\'o}rski} et~al.}{2005}]{healpix}
{G{\'o}rski} K.~M.,  {Hivon} E.,  {Banday} A.~J.,  {Wandelt} B.~D.,  {Hansen}
  F.~K.,  {Reinecke} M.,    {Bartelmann} M.,  2005, \apj, 622, 759

\bibitem[\protect\citeauthoryear{{Grenier}, {Casandjian} \&
  {Terrier}}{{Grenier} et~al.}{2005}]{Grenier05}
{Grenier} I.~A.,  {Casandjian} J.-M.,    {Terrier} R.,  2005, Science, 307,
  1292

\bibitem[\protect\citeauthoryear{{Habing}}{{Habing}}{1968}]{Habing68}
{Habing} H.~J.,  1968, \bain, 19, 421

\bibitem[\protect\citeauthoryear{{Henning}, {Linz}, {Krause}, {Ragan},
  {Beuther}, {Launhardt}, {Nielbock} \& {Vasyunina}}{{Henning}
  et~al.}{2010}]{Henning10}
{Henning} T.,  {Linz} H.,  {Krause} O.,  {Ragan} S.,  {Beuther} H.,
  {Launhardt} R.,  {Nielbock} M.,    {Vasyunina} T.,  2010, \aap, 518, L95

\bibitem[\protect\citeauthoryear{{Hopkins}, {Quataert} \& {Murray}}{{Hopkins}
  et~al.}{2012}]{Hopkins12a}
{Hopkins} P.~F.,  {Quataert} E.,    {Murray} N.,  2012, \mnras, 421, 3488

\bibitem[\protect\citeauthoryear{{Jackson}, {Finn}, {Chambers}, {Rathborne} \&
  {Simon}}{{Jackson} et~al.}{2010}]{Jackson10}
{Jackson} J.~M.,  {Finn} S.~C.,  {Chambers} E.~T.,  {Rathborne} J.~M.,
  {Simon} R.,  2010, \apjl, 719, L185

\bibitem[\protect\citeauthoryear{{Langer}, {Velusamy}, {Pineda}, {Goldsmith},
  {Li} \& {Yorke}}{{Langer} et~al.}{2010}]{Langer10}
{Langer} W.~D.,  {Velusamy} T.,  {Pineda} J.~L.,  {Goldsmith} P.~F.,  {Li} D.,
    {Yorke} H.~W.,  2010, \aap, 521, L17

\bibitem[\protect\citeauthoryear{{Langer}, {Velusamy}, {Pineda}, {Willacy} \&
  {Goldsmith}}{{Langer} et~al.}{2014}]{Langer14}
{Langer} W.~D.,  {Velusamy} T.,  {Pineda} J.~L.,  {Willacy} K.,    {Goldsmith}
  P.~F.,  2014, \aap, 561, A122

\bibitem[\protect\citeauthoryear{{Lee}, {Herbst}, {Pineau des Forets}, {Roueff}
  \& {Le Bourlot}}{{Lee} et~al.}{1996}]{Lee96}
{Lee} H.-H.,  {Herbst} E.,  {Pineau des Forets} G.,  {Roueff} E.,    {Le
  Bourlot} J.,  1996, \aap, 311, 690

\bibitem[\protect\citeauthoryear{{Lee}, {Stanimirovic}, {Wolfire}, {Shetty},
  {Glover}, {Molina} \& {Klessen}}{{Lee} et~al.}{2014}]{Lee2014}
{Lee} M.-Y.,  {Stanimirovic} S.,  {Wolfire} M.~G.,  {Shetty} R.,  {Glover}
  S.~C.~O.,  {Molina} F.~Z.,    {Klessen} R.~S.,  2014, \apj, 784, 80

\bibitem[\protect\citeauthoryear{{Leroy}, {Bolatto}, {Stanimirovic}, {Mizuno},
  {Israel} \& {Bot}}{{Leroy} et~al.}{2007}]{Leroy07}
{Leroy} A.,  {Bolatto} A.,  {Stanimirovic} S.,  {Mizuno} N.,  {Israel} F.,
  {Bot} C.,  2007, \apj, 658, 1027

\bibitem[\protect\citeauthoryear{{Li}, {Wyrowski}, {Menten} \& {Belloche}}{{Li}
  et~al.}{2013}]{Li13}
{Li} G.-X.,  {Wyrowski} F.,  {Menten} K.,    {Belloche} A.,  2013, \aap, 559,
  A34

\bibitem[\protect\citeauthoryear{{Liszt} \& {Pety}}{{Liszt} \&
  {Pety}}{2012}]{Liszt12}
{Liszt} H.~S.,  {Pety} J.,  2012, \aap, 541, A58

\bibitem[\protect\citeauthoryear{{Luck}, {Andrievsky}, {Kovtyukh}, {Gieren} \&
  {Graczyk}}{{Luck} et~al.}{2011}]{Luck11}
{Luck} R.~E.,  {Andrievsky} S.~M.,  {Kovtyukh} V.~V.,  {Gieren} W.,
  {Graczyk} D.,  2011, \aj, 142, 51

\bibitem[\protect\citeauthoryear{{Ma{\'{\i}}z-Apell{\'a}niz}}{{Ma{\'{\i}}z-Apell{\'a}niz}}{2001}]{Maiz-Apellaniz01}
{Ma{\'{\i}}z-Apell{\'a}niz} J.,  2001, \aj, 121, 2737

\bibitem[\protect\citeauthoryear{{Nelson} \& {Langer}}{{Nelson} \&
  {Langer}}{1997}]{Nelson97}
{Nelson} R.~P.,  {Langer} W.~D.,  1997, \apj, 482, 796

\bibitem[\protect\citeauthoryear{{Neufeld} et~al.,}{{Neufeld}
  et~al.}{2010}]{Neufeld10}
{Neufeld} D.~A.,  et~al., 2010, \aap, 521, L10

\bibitem[\protect\citeauthoryear{{Pakmor}, {Bauer} \& {Springel}}{{Pakmor}
  et~al.}{2011}]{Pakmor11}
{Pakmor} R.,  {Bauer} A.,    {Springel} V.,  2011, \mnras, 418, 1392

\bibitem[\protect\citeauthoryear{{Paradis}, {Dobashi}, {Shimoikura},
  {Kawamura}, {Onishi}, {Fukui} \& {Bernard}}{{Paradis}
  et~al.}{2012}]{Paradis12}
{Paradis} D.,  {Dobashi} K.,  {Shimoikura} T.,  {Kawamura} A.,  {Onishi} T.,
  {Fukui} Y.,    {Bernard} J.-P.,  2012, \aap, 543, A103

\bibitem[\protect\citeauthoryear{{Pilbratt}, {Riedinger}, {Passvogel}, {Crone},
  {Doyle}, {Gageur}, {Heras}, {Jewell}, {Metcalfe}, {Ott} \&
  {Schmidt}}{{Pilbratt} et~al.}{2010}]{Pilbratt10}
{Pilbratt} G.~L.,  {Riedinger} J.~R.,  {Passvogel} T.,  {Crone} G.,  {Doyle}
  D.,  {Gageur} U.,  {Heras} A.~M.,  {Jewell} C.,  {Metcalfe} L.,  {Ott} S.,
  {Schmidt} M.,  2010, \aap, 518, L1

\bibitem[\protect\citeauthoryear{{Pineda}, {Caselli} \& {Goodman}}{{Pineda}
  et~al.}{2008}]{Pineda08}
{Pineda} J.~E.,  {Caselli} P.,    {Goodman} A.~A.,  2008, \apj, 679, 481

\bibitem[\protect\citeauthoryear{{Pineda}, {Langer}, {Velusamy} \&
  {Goldsmith}}{{Pineda} et~al.}{2013}]{PinedaJ13}
{Pineda} J.~L.,  {Langer} W.~D.,  {Velusamy} T.,    {Goldsmith} P.~F.,  2013,
  \aap, 554, A103

\bibitem[\protect\citeauthoryear{{Planck Collaboration}, {Ade}, {Aghanim},
  {Arnaud}, {Ashdown}, {Aumont}, {Baccigalupi}, {Balbi}, {Banday}, {Barreiro}
  \& et al.}{{Planck Collaboration} et~al.}{2011}]{Planck11}
{Planck Collaboration} {Ade} P.~A.~R.,  {Aghanim} N.,  {Arnaud} M.,  {Ashdown}
  M.,  {Aumont} J.,  {Baccigalupi} C.,  {Balbi} A.,  {Banday} A.~J.,
  {Barreiro} R.~B.,    et al. 2011, \aap, 536, A19

\bibitem[\protect\citeauthoryear{{Ragan}, {Henning}, {Tackenberg}, {Beuther},
  {Johnston}, {Kainulainen} \& {Linz}}{{Ragan} et~al.}{2014}]{Ragan14}
{Ragan} S.~E.,  {Henning} T.,  {Tackenberg} J.,  {Beuther} H.,  {Johnston}
  K.~G.,  {Kainulainen} J.,    {Linz} H.,  2014, A\&A, submitted;
  arXiv:1403.1450

\bibitem[\protect\citeauthoryear{{Reed}}{{Reed}}{2000}]{Reed00}
{Reed} B.~C.,  2000, \aj, 120, 314

\bibitem[\protect\citeauthoryear{{Roman-Duval} et~al.,}{{Roman-Duval}
  et~al.}{2010}]{RomanDuval10}
{Roman-Duval} J.,  et~al., 2010, \aap, 518, L74

\bibitem[\protect\citeauthoryear{{Savage}, {Bohlin}, {Drake} \&
  {Budich}}{{Savage} et~al.}{1977}]{Savage77}
{Savage} B.~D.,  {Bohlin} R.~C.,  {Drake} J.~F.,    {Budich} W.,  1977, \apj,
  216, 291

\bibitem[\protect\citeauthoryear{{Schinnerer}, {Meidt}, {Pety}, {Hughes},
  {Colombo}, {Garc{\'{\i}}a-Burillo}, {Schuster}, {Dumas}, {Dobbs}, {Leroy},
  {Kramer}, {Thompson} \& {Regan}}{{Schinnerer} et~al.}{2013}]{Schinnerer13}
{Schinnerer} E.,  {Meidt} S.~E.,  {Pety} J.,  {Hughes} A.,  {Colombo} D.,
  {Garc{\'{\i}}a-Burillo} S.,  {Schuster} K.~F.,  {Dumas} G.,  {Dobbs} C.~L.,
  {Leroy} A.~K.,  {Kramer} C.,  {Thompson} T.~A.,    {Regan} M.~W.,  2013,
  \apj, 779, 42

\bibitem[\protect\citeauthoryear{{Sch{\"o}ier}, {van der Tak}, {van Dishoeck}
  \& {Black}}{{Sch{\"o}ier} et~al.}{2005}]{Schoier05}
{Sch{\"o}ier} F.~L.,  {van der Tak} F.~F.~S.,  {van Dishoeck} E.~F.,    {Black}
  J.~H.,  2005, \aap, 432, 369

\bibitem[\protect\citeauthoryear{{Sheffer}, {Rogers}, {Federman}, {Abel},
  {Gredel}, {Lambert} \& {Shaw}}{{Sheffer} et~al.}{2008}]{Sheffer08}
{Sheffer} Y.,  {Rogers} M.,  {Federman} S.~R.,  {Abel} N.~P.,  {Gredel} R.,
  {Lambert} D.~L.,    {Shaw} G.,  2008, \apj, 687, 1075

\bibitem[\protect\citeauthoryear{{Shetty}, {Glover}, {Dullemond}, {Ostriker},
  {Harris} \& {Klessen}}{{Shetty} et~al.}{011b}]{Shetty11b}
{Shetty} R.,  {Glover} S.~C.,  {Dullemond} C.~P.,  {Ostriker} E.~C.,  {Harris}
  A.~I.,    {Klessen} R.~S.,  2011b, \mnras, 415, 3253

\bibitem[\protect\citeauthoryear{{Sijacki}, {Vogelsberger}, {Kere{\v s}},
  {Springel} \& {Hernquist}}{{Sijacki} et~al.}{2012}]{Sijacki12}
{Sijacki} D.,  {Vogelsberger} M.,  {Kere{\v s}} D.,  {Springel} V.,
  {Hernquist} L.,  2012, \mnras, 424, 2999

\bibitem[\protect\citeauthoryear{{Snow} \& {McCall}}{{Snow} \&
  {McCall}}{2006}]{Snow06}
{Snow} T.~P.,  {McCall} B.~J.,  2006, \araa, 44, 367

\bibitem[\protect\citeauthoryear{{Sonnentrucker}, {Welty}, {Thorburn} \&
  {York}}{{Sonnentrucker} et~al.}{2007}]{Sonnentrucker07}
{Sonnentrucker} P.,  {Welty} D.~E.,  {Thorburn} J.~A.,    {York} D.~G.,  2007,
  \apjs, 168, 58

\bibitem[\protect\citeauthoryear{{Springel}}{{Springel}}{2010}]{Springel10}
{Springel} V.,  2010, \mnras, 401, 791

\bibitem[\protect\citeauthoryear{{Tackenberg}, {Beuther}, {Plume}, {Henning},
  {Stil}, {Walmsley}, {Schuller} \& {Schmiedeke}}{{Tackenberg}
  et~al.}{2013}]{Tackenberg13}
{Tackenberg} J.,  {Beuther} H.,  {Plume} R.,  {Henning} T.,  {Stil} J.,
  {Walmsley} M.,  {Schuller} F.,    {Schmiedeke} A.,  2013, \aap, 550, A116

\bibitem[\protect\citeauthoryear{{Tielens}}{{Tielens}}{2005}]{Tielens05}
{Tielens} A.~G.~G.~M.,  2005, {The Physics and Chemistry of the Interstellar
  Medium}.
Cambridge University Press, Cambridge

\bibitem[\protect\citeauthoryear{{Tielens} \& {Hollenbach}}{{Tielens} \&
  {Hollenbach}}{1985}]{Tielens85}
{Tielens} A.~G.~G.~M.,  {Hollenbach} D.,  1985, \apj, 291, 722

\bibitem[\protect\citeauthoryear{{van der Tak} \& {van Dishoeck}}{{van der Tak}
  \& {van Dishoeck}}{2000}]{vv2000}
{van der Tak} F.~F.~S.,  {van Dishoeck} E.~F.,  2000, \aap, 358, L79

\bibitem[\protect\citeauthoryear{{van Dishoeck} \& {Black}}{{van Dishoeck} \&
  {Black}}{1988}]{vanDishoeck88}
{van Dishoeck} E.~F.,  {Black} J.~H.,  1988, \apj, 334, 771

\bibitem[\protect\citeauthoryear{{Wolfire}, {Hollenbach} \& {McKee}}{{Wolfire}
  et~al.}{2010}]{Wolfire10}
{Wolfire} M.~G.,  {Hollenbach} D.,    {McKee} C.~F.,  2010, \apj, 716, 1191

\bibitem[\protect\citeauthoryear{{Wolfire}, {McKee}, {Hollenbach} \&
  {Tielens}}{{Wolfire} et~al.}{2003}]{Wolfire03}
{Wolfire} M.~G.,  {McKee} C.~F.,  {Hollenbach} D.,    {Tielens} A.~G.~G.~M.,
  2003, \apj, 587, 278

\bibitem[\protect\citeauthoryear{{Wolfire}, {Tielens}, {Hollenbach} \&
  {Kaufman}}{{Wolfire} et~al.}{2008}]{Wolfire08}
{Wolfire} M.~G.,  {Tielens} A.~G.~G.~M.,  {Hollenbach} D.,    {Kaufman} M.~J.,
  2008, \apj, 680, 384

\bibitem[\protect\citeauthoryear{{Zsarg{\'o}} \& {Federman}}{{Zsarg{\'o}} \&
  {Federman}}{2003}]{Zsargo03}
{Zsarg{\'o}} J.,  {Federman} S.~R.,  2003, \apj, 589, 319

\end{thebibliography}
\label{lastpage}

\end{document}